\documentclass[prd,aps,floats,twocolumn,nofootinbib,superscriptaddress]{revtex4-2}
\usepackage{natbib}
\usepackage{commath}
\pdfoutput=1
\usepackage[usenames, dvipsnames]{color}
\usepackage[T1]{fontenc}
\usepackage{lmodern}
\usepackage{graphicx, array}
\usepackage{multirow}
\usepackage{graphics}
\usepackage{graphics,epsfig}
\usepackage{amssymb}
\usepackage{amsmath}
\usepackage{ulem}
\usepackage{multirow}
\usepackage{subfigure}
\usepackage{bm}
\usepackage{txfonts}
\usepackage{cancel}
\usepackage{url}
\usepackage{hyperref}
\usepackage{lipsum}
\usepackage{graphicx}
\usepackage{comment} 
\usepackage{soul}

\newcolumntype{C}[1]{>{\centering\arraybackslash}m{#1}}

\def\be{\begin{equation}}
\def\ee{\end{equation}}
\def\bi{\begin{itemize}}
\def\ei{\end{itemize}}
\def\ben{\begin{enumerate}}
\def\een{\end{enumerate}}
\def\bt{\begin{tabular}}
\def\et{\end{tabular}}
\def\bc{\begin{center}}
\def\ec{\end{center}}

\def\bea{\begin{eqnarray}}
\def\eea{\end{eqnarray}}

\def\ba{\begin{eqnarray}}
\def\ea{\end{eqnarray}}

\let\oldhat\hat

\renewcommand{\hat}[1]{\oldhat{\boldsymbol{\mathbf{#1}}}}

\graphicspath{ {./images/},{ ./images/ratioFigs/},{ ./images/windowFigs/},{ ./Version1/}}

\begin{document}

\title{Detection of the Pairwise Kinematic Sunyaev-Zel’dovich Eﬀect and Pairwise Velocity with DESI DR1 Galaxies and ACT DR6 and \textit{Planck} CMB data}

\author{Yulin Gong}
\affiliation{Department of Astronomy, Cornell University, Ithaca, NY 14853, USA}
\author{Patricio A. Gallardo} 
\affiliation{Department of Physics and Astronomy, University of Pennsylvania, 209 South 33rd Street, Philadelphia, PA, USA 19104}
\author{Rachel Bean}
\affiliation{Department of Astronomy, Cornell University, Ithaca, NY 14853, USA}
\author{Jenna Moore}
\author{Eve M. Vavagiakis}
\affiliation{Department of Physics, Duke University, Durham, NC, 27708}
\author{Nicholas Battaglia}
\affiliation{Department of Astronomy, Cornell University, Ithaca, NY 14853, USA}
\author{Boryana Hadzhiyska}
\affiliation{Institute of Astronomy, Madingley Road, Cambridge, CB3 0HA, UK}
\affiliation{Kavli Institute for Cosmology Cambridge, Madingley Road, Cambridge, CB3 0HA, UK}
\author{Yun-Hsin Hsu}
\affiliation{University Observatory, Faculty of Physics, Ludwig-Maximilians-Universität, Scheinerstr. 1, 81679 Munich, Germany}

\author{Jessica Nicole Aguilar}
\affiliation{Lawrence Berkeley National Laboratory, 1 Cyclotron Road, Berkeley, CA 94720, USA}
\author{Steven Ahlen}
\affiliation{Department of Physics, Boston University, 590 Commonwealth Avenue, Boston, MA 02215 USA}
\author{Davide Bianchi}
\affiliation{Dipartimento di Fisica ``Aldo Pontremoli'', Universit\`a degli Studi di Milano, Via Celoria 16, I-20133 Milano, Italy}
\affiliation{INAF-Osservatorio Astronomico di Brera, Via Brera 28, 20122 Milano, Italy}
\author{David Brooks}
\affiliation{Department of Physics \& Astronomy, University College London, Gower Street, London, WC1E 6BT, UK}
\author{Todd Claybaugh}
\affiliation{Lawrence Berkeley National Laboratory, 1 Cyclotron Road, Berkeley, CA 94720, USA}
\author{Rebecca Canning}
\affiliation{Institute of Cosmology and Gravitation, University of Portsmouth, Dennis Sciama Building, Portsmouth, PO1 3FX, UK}

\author{Mark Devlin}
\affiliation{Department of Physics and Astronomy, University of Pennsylvania, 209 South 33rd Street, Philadelphia, PA, USA 19104}

\author{Peter Doel}
\affiliation{Department of Physics \& Astronomy, University College London, Gower Street, London, WC1E 6BT, UK}
\author{Axel de la Macorra}
\affiliation{Instituto de F\'{\i}sica, Universidad Nacional Aut\'{o}noma de M\'{e}xico,  Circuito de la Investigaci\'{o}n Cient\'{\i}fica, Ciudad Univ
ersitaria, Cd. de M\'{e}xico  C.~P.~04510,  M\'{e}xico}
\author{Simone Ferraro}
\affiliation{Lawrence Berkeley National Laboratory, 1 Cyclotron Road, Berkeley, CA 94720, USA}
\affiliation{University of California, Berkeley, 110 Sproul Hall \#5800 Berkeley, CA 94720, USA}
\author{Andreu Font-Ribera}
\affiliation{Institut de F\'{i}sica d’Altes Energies (IFAE), The Barcelona Institute of Science and Technology, Edifici Cn, Campus UAB, 08193, Bellate
rra (Barcelona), Spain}
\author{Jaime E. Forero-Romero}
\affiliation{Departamento de F\'isica, Universidad de los Andes, Cra. 1 No. 18A-10, Edificio Ip, CP 111711, Bogot\'a, Colombia}
\affiliation{Observatorio Astron\'omico, Universidad de los Andes, Cra. 1 No. 18A-10, Edificio H, CP 111711 Bogot\'a, Colombia}
\author{Enrique Gaztañaga}
\affiliation{Institut d'Estudis Espacials de Catalunya (IEEC), c/ Esteve Terradas 1, Edifici RDIT, Campus PMT-UPC, 08860 Castelldefels, Spain}
\affiliation{Institute of Space Sciences, ICE-CSIC, Campus UAB, Carrer de Can Magrans s/n, 08913 Bellaterra, Barcelona, Spain}
\affiliation{Institute of Cosmology and Gravitation, University of Portsmouth, Dennis Sciama Building, Portsmouth, PO1 3FX, UK}
\author{Gaston Gutierrez}
\affiliation{Fermi National Accelerator Laboratory, PO Box 500, Batavia, IL 60510, USA}

\author{Satya Gontcho A Gontcho}
\affiliation{Lawrence Berkeley National Laboratory, 1 Cyclotron Road, Berkeley, CA 94720, USA}
\affiliation{University of Virginia, Department of Astronomy, Charlottesville, VA 22904, USA}

\author{Julien Guy}
\affiliation{Lawrence Berkeley National Laboratory, 1 Cyclotron Road, Berkeley, CA 94720, USA}

\author{Klaus Honscheid}
\affiliation{Center for Cosmology and AstroParticle Physics, The Ohio State University, 191 West Woodruff Avenue, Columbus, OH 43210, USA}
\affiliation{Department of Physics, The Ohio State University, 191 West Woodruff Avenue, Columbus, OH 43210, USA}
\affiliation{The Ohio State University, Columbus, 43210 OH, USA}

\author{Cullan Howlett}
\affiliation{School of Mathematics and Physics, University of Queensland, Brisbane, QLD 4072, Australia}
\author{R. Henry Liu}
\affiliation{Department of Physics, University of California, Berkeley, CA, 94720, USA}
\affiliation{Lawrence Berkeley National Laboratory, 1 Cyclotron Road, Berkeley, CA 94720, USA}

\author{Mustapha Ishak}
\affiliation{Department of Physics, The University of Texas at Dallas, 800 W. Campbell Rd., Richardson, TX 75080, USA}

\author{Dick Joyce}
\affiliation{NSF NOIRLab, 950 N. Cherry Ave., Tucson, AZ 85719, USA}

\author{Anthony Kremin}
\affiliation{Lawrence Berkeley National Laboratory, 1 Cyclotron Road, Berkeley, CA 94720, USA}
\author{Claire Lamman}
\affiliation{The Ohio State University, Columbus, 43210 OH, USA}
\author{Michael Levi}
\affiliation{Lawrence Berkeley National Laboratory, 1 Cyclotron Road, Berkeley, CA 94720, USA}
\author{Martin Landriau}
\affiliation{Lawrence Berkeley National Laboratory, 1 Cyclotron Road, Berkeley, CA 94720, USA}

\author{Marc Manera}
\affiliation{Departament de F\'{i}sica, Serra H\'{u}nter, Universitat Aut\`{o}noma de Barcelona, 08193 Bellaterra (Barcelona), Spain}
\affiliation{Institut de F\'{i}sica d‚ÄôAltes Energies (IFAE), The Barcelona Institute of Science and Technology, Edifici Cn, Campus UAB, 08193, Bellaterra (Barcelona), Spain}
\author{Aaron Meisner}
\affiliation{NSF NOIRLab, 950 N. Cherry Ave., Tucson, AZ 85719, USA}
\author{Ramon Miquel}
\affiliation{Instituci\'{o} Catalana de Recerca i Estudis Avancats, Passeig de Llu\'{\i}s Companys, 23, 08010 Barcelona, Spain}
\affiliation{Institut de F\'{i}sica d’Altes Energies (IFAE), The Barcelona Institute of Science and Technology, Edifici Cn, Campus UAB, 08193, Bellate
rra (Barcelona), Spain}
\author{Michael D. Niemack}
\affiliation{Department of Physics, Cornell University, Ithaca, NY 14853,USA}
\affiliation{Department of Astronomy, Cornell University, Ithaca, NY 14853, USA}
\author{Seshadri Nadathur}
\affiliation{Institute of Cosmology and Gravitation, University of Portsmouth, Dennis Sciama Building, Portsmouth, PO1 3FX, UK}
\author{Will Percival}
\affiliation{Department of Physics and Astronomy, University of Waterloo, 200 University Ave W, Waterloo, ON N2L 3G1, Canada}
\affiliation{Perimeter Institute for Theoretical Physics, 31 Caroline St. North, Waterloo, ON N2L 2Y5, Canada}
\affiliation{Waterloo Centre for Astrophysics, University of Waterloo, 200 University Ave W, Waterloo, ON N2L 3G1, Canada}
\author{Francisco Prada}
\affiliation{Instituto de Astrof\'{i}sica de Andaluc\'{i}a (CSIC), Glorieta de la Astronom\'{i}a, s/n, E-18008 Granada, Spain}
\author{Graziano Rossi}
\affiliation{Department of Physics and Astronomy, Sejong University, 209 Neungdong-ro, Gwangjin-gu, Seoul 05006, Republic of Korea}
\author{Bernardita Ried Guachalla}
\affiliation{Department of Physics, Stanford University, Stanford, CA, USA 94305-4085}
\affiliation{Kavli Institute for Particle Astrophysics and Cosmology, 382 Via Pueblo Mall Stanford, CA 94305-4060, USA}
\affiliation{SLAC National Accelerator Laboratory 2575 Sand Hill Road Menlo Park, California 94025, USA} 
\author{Eusebio Sanchez}
\affiliation{CIEMAT, Avenida Complutense 40, E-28040 Madrid, Spain}
\author{Hee-Jong Seo}
\affiliation{Department of Physics \& Astronomy, Ohio University, 139 University Terrace, Athens, OH 45701, USA}
\author{David Sprayberry}
\affiliation{NSF NOIRLab, 950 N. Cherry Ave., Tucson, AZ 85719, USA}

\author{David Schlegel}
\affiliation{Lawrence Berkeley National Laboratory, 1 Cyclotron Road, Berkeley, CA 94720, USA}

\author{Crist\'obal Sif\'on}
\affiliation{Instituto de F\'isica, Pontificia Universidad Cat\'olica de Valpara\'iso, Casilla 4059, Valpara\'iso, Chile}

\author{Michael Schubnell}
\affiliation{Department of Physics, University of Michigan, 450 Church Street, Ann Arbor, MI 48109, USA}
\affiliation{University of Michigan, 500 S. State Street, Ann Arbor, MI 48109, USA}
\author{Joseph Harry Silber}
\affiliation{Lawrence Berkeley National Laboratory, 1 Cyclotron Road, Berkeley, CA 94720, USA}

\author{Gregory Tarl\'{e}}
\affiliation{University of Michigan, 500 S. State Street, Ann Arbor, MI 48109, USA}
\author{Benjamin Alan Weaver}
\affiliation{NSF NOIRLab, 950 N. Cherry Ave., Tucson, AZ 85719, USA}

\author{Rongpu Zhou}
\affiliation{Lawrence Berkeley National Laboratory, 1 Cyclotron Road, Berkeley, CA 94720, USA}
\author{Hu Zou}
\affiliation{National Astronomical Observatories, Chinese Academy of Sciences, A20 Datun Road, Chaoyang District, Beijing, 100101, P.~R.~China}

\date{\today}

\begin{abstract}
We present a 9.3$\sigma$ detection of the pairwise kinematic Sunyaev-Zel’dovich (kSZ) effect by combining a sample of 913,286 Luminous Red Galaxies (LRGs) from the Dark Energy Spectroscopic Instrument Data Release 1 (DESI DR1) catalog and co-added Atacama Cosmology Telescope (ACT DR6) and \textit{Planck} cosmic microwave background (CMB) temperature maps. This represents the highest-significance pairwise kSZ measurement to date. The analysis uses three  ACT CMB temperature maps: co-added 150 GHz, total frequency maps, and a component separated Internal Linear Combination (ILC) map, all of which cover 19,000 square degrees of the sky from Advanced ACTPol observations conducted between 2017 and 2022. Comparison of the results of these three maps serves as a consistency check for potential foreground contamination that may depend on the observation frequency. An estimate of the best-fit mass-averaged optical depth is obtained by comparing the pairwise kSZ curve with the linear theory prediction of the pairwise velocity under the best-fit \textit{Planck} cosmology and is compared with predictions from simulations. This estimate serves as a reference point for future comparisons with thermal SZ–derived optical depth measurements for the same DESI cluster samples, which will be presented in a companion paper. Finally, we employ a machine learning approach, trained on simulations to estimate the optical depth for 456,803 DESI LRG-identified clusters within the simulated mass range ($ \gtrsim 10^{13}M_{\odot}$). These are combined with the measured kSZ signal to infer the individual cluster peculiar velocities, providing the opportunity to constrain the behavior of gravity and the dark sector over a range of cosmic scales and epochs.
\end{abstract}

\maketitle
 
\section{Introduction}
\label{sec:intro}
The kinetic Sunyaev-Zel'dovich (kSZ) effect, which arises from the line-of-sight peculiar motion of free electrons in galaxy clusters relative to the Cosmic Microwave Background (CMB) rest frame, serves as a powerful probe of cosmological structure formation and the large-scale velocity field of the Universe \cite{Sunyaev:1980nv,Sunyaev:1970bma}. The detection of the kSZ effect also provides an independent probe of large-scale structure to redshift-space distortions (RSD) \cite{Hamilton:1997zq,Marulli:2015jga,Okumura:2015lvp,Tonegawa:2020wuh,Eggemeier:2025xwi} and weak gravitational lensing \cite{Refregier:2003ct, Hoekstra:2008db, Kilbinger:2014cea, Wenzl:2024sug, Prat:2025ucy}. In contrast with the thermal Sunyaev-Zel'dovich (tSZ) effect, which originates from inverse Compton scattering of CMB photons by hot electrons in clusters, the kSZ effect directly traces the peculiar velocities of galaxy clusters and groups. This complements alternative methods with analyses sensitive to galaxy bias and nonlinear clustering and weak lensing requiring precise shape measurements of background galaxies, making it a valuable tool for studying cosmological structures and offering an independent avenue for testing models of modified gravity and alternative dark energy models \cite{Clifton:2011jh,Joyce:2014kja,Joyce:2016vqv,Nojiri:2017ncd,Langlois:2018dxi}.

The pairwise kSZ statistic, which measures the average relative momentum of galaxy or cluster pairs, has emerged as an effective approach to detect the kSZ signal \cite{1983ApJ...267..465D}. Using the large-scale coherence of velocity fields, this method suppresses uncorrelated noise and significantly enhances detection sensitivity compared to single-object velocity measurements. Through statistical analysis of cluster or galaxy pairs, it becomes possible to probe cosmic velocity fields and constrain key cosmological parameters, including the amplitude of matter fluctuations, the growth rate of structure, neutrino masses, and the nature of dark energy \cite{DeDeo:2005yr,Bhattacharya:2007sk,Kosowsky_2009,Bull:2011wi,Mueller:2014dba,Mueller:2014nsa}.

Luminous Red Galaxies (LRGs) are an important and useful tool for many scientific studies. Since the most luminous galaxies within clusters are typically red and bright \citep{1972ApJ...178....1S,1980ApJ...241..486H,1983ApJ...264..337S,1995ApJ...440...28P}, the main scientific goal of the LRG sample is to identify and trace clusters of galaxies to study large-scale structure \cite{SDSS:2001wju,Zheng:2008np}. The first detections of the pairwise kSZ effect were obtained by combining Atacama Cosmology Telescope (ACT) data with LRG samples from the Sloan Digital Sky Survey (SDSS), providing the first observational evidence of the kSZ signal and marking a key milestone in cosmology \cite{Hand_2012}. This measurement demonstrated the feasibility of using the pairwise kSZ signal as a robust cosmological probe, motivating subsequent analyses with increasingly sophisticated techniques and larger datasets. Building on this foundation, later studies have extracted the signal with higher precision by combining improved galaxy catalogs from BOSS, DES, and DESI with later CMB maps from ACT and \textit{Planck}, leading to higher sensitivity measurements and opening the door to cosmological constraints \cite{DeBernardis:2016pdv,Planck:2015ywj,DES:2016umt,Calafut:2021wkx,Kusiak:2021hai,Chen:2021pwg,Li:2024svf}.

The Dark Energy Spectroscopic Instrument (DESI) survey \cite{DESI:2016fyo} is a next-generation galaxy redshift survey designed to map the large-scale structure of the Universe with unprecedented precision. DESI provides high-precision spectroscopic redshifts for tens of millions of galaxies, quasars, and other cosmological tracers across a wide redshift range. Meanwhile, ACT Data Release 6 (DR6) delivers the deepest ground-based CMB maps to date, with improved sensitivity, wide sky coverage, and broad frequency coverage \cite{Naess:2020wgi}. Coupled with the DESI Data Release 1 (DR1) galaxy catalog, which contains precise spectroscopic redshifts for millions of galaxies, this analysis significantly improves the sensitivity of the LRG kSZ measurements.

In this work, we exploit the synergy between ACT DR6 maps and the DESI DR1 galaxy catalog to perform a comprehensive analysis of the pairwise kSZ signal. Our goal is to sensitively measure the pairwise kSZ effect and address the limitations of previous LRG studies. In addition, we incorporate machine learning methods to directly infer individual cluster optical depths and peculiar velocities, thereby enabling more robust pairwise velocity estimates. This study forms part of an upcoming series of analyses leveraging DESI DR1 and ACT DR6 data to measure the pairwise kSZ and tSZ effects with various galaxy and cluster tracers \citep[in prep.]{Hadzhiyska:2025egz,Hsu2025,Moore2025}.

The structure of this paper is as follows. Section \ref{sec:data} describes the datasets used in our analysis, focusing on the characteristics of ACT DR6 and DESI DR1. Section \ref{sec:methods} introduces the methodology and statistical tools. Section \ref{sec:results} presents the main results, and Section \ref{sec:conclusion} summarizes the key findings and discusses the prospects for future improvements and analyses.

\section{Data}
\label{sec:data}

\subsection{ACT DR6 data}
The Atacama Cosmology Telescope (ACT) was a ground-based millimeter-wavelength telescope that was designed to map the CMB with high angular resolution and sensitivity, allowing detailed measurements of temperature and polarization anisotropies across a wide range of angular scales \cite{Swetz_2011,Thornton_2016}. The resulting dataset enabled high-precision studies of CMB power spectra, gravitational lensing of the CMB, and the SZ effect from massive galaxy clusters, providing independent constraints on the standard cosmological model and the history of structure formation in the Universe \cite{ACT:2020frw}. This analysis uses three CMB temperature maps from ACT Data Release 6 (DR6) \cite{ACT:2025xdm}. The maps are derived from data collected between 2017 and 2022 and are combined with \textit{Planck} data \cite{Lamarre:2003zh}, covering a sky area of 19,000 square degrees. 

The ``f150" map is a combined frequency-specific map centered at 150 GHz (124–172 GHz). The ``ftot" map is an optimally inverse-noise-weighted combination of observations at 98, 150, and 220 GHz, designed to maximize the signal-to-noise ratio for CMB anisotropies. Finally, we use the DR6 CMB internal linear combination (``ILC") map \cite{ACT:2024rue} without tSZ deprojection, which integrates multi-frequency data from ACT and complementary information from \textit{Planck}, spanning frequencies from 30 GHz to 545 GHz. 

In addition to the CMB ILC temperature maps, we also use the DR6 Compton-y ILC map \cite{ACT:2023wcq}, which provides a measurement of the thermal tSZ effect across the sky. In this case, the ILC technique isolates the tSZ signal through linearly combining maps at different frequencies in a way that minimizes contamination from other astrophysical foregrounds and instrumental noise, while preserving the unique spectral signature of the tSZ effect. ACT DR6 CMB gravitational lensing convergence maps, covering the same sky area as the three CMB temperature maps described above, are also used in this work \cite{ACT:2023dou,ACT:2023kun}. These trace the projected total matter density along the line of sight, reconstructed from the subtle distortions imprinted on the CMB by intervening large-scale structures. These maps are not used directly in the pairwise kSZ–estimator analysis, but instead are employed to train the models used to infer the optical depth.

\subsection{DESI DR1 data}

\begin{figure}
    \includegraphics[width=\columnwidth]{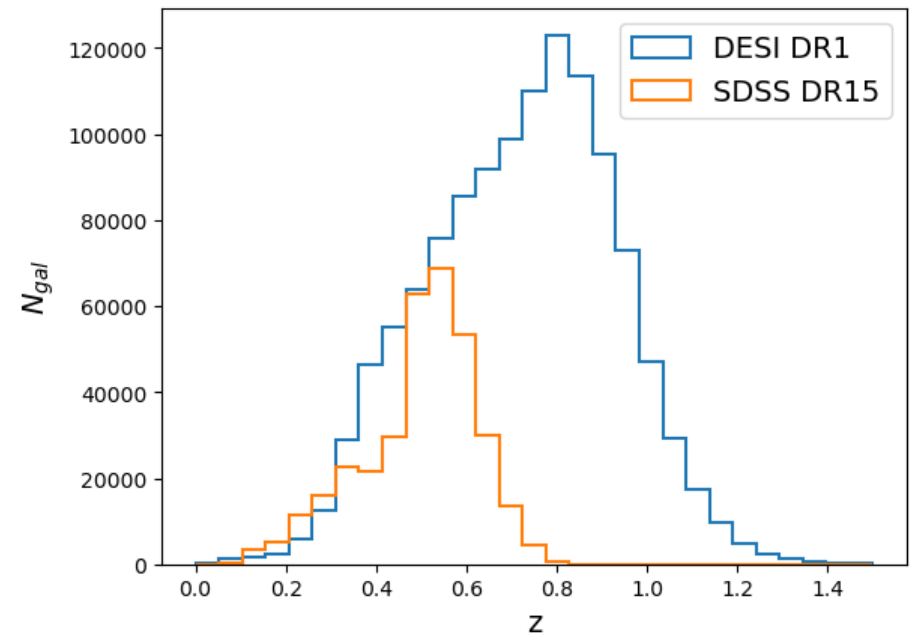}
    \caption{The redshift distribution, showing the number of LRGs, $N_{gal}$, in the sample from DESI DR1  [blue] analyzed in this work and, as a reference, from SDSS DR15 [orange] studied previously in \cite{Calafut:2021wkx}.}
    \label{fig:redshift}
\end{figure}

\begin{figure*}
    \includegraphics[width=
    \textwidth]{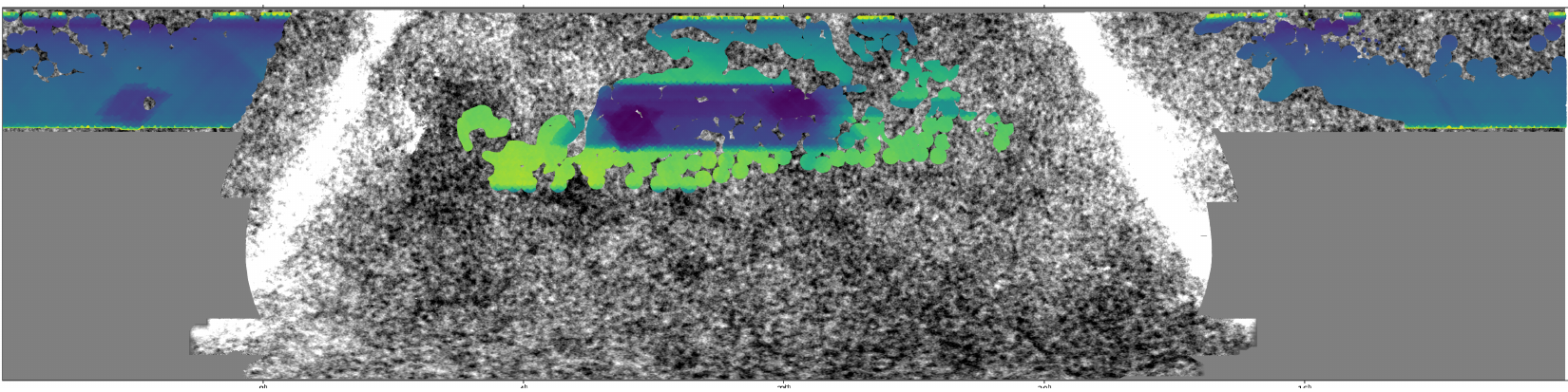}
    \caption{ACT DR6 and DESI sky coverage. Background shows the ACT DR6 150GHz map. Shaded colored region shows the selected DESI region footprint, and color shows the average noise level in the ACT 150 GHz map. Gaps in coverage result of combining the DESI footprint with the point source, and galactic mask as described in Section~\ref{sec:data}.}
    \label{fig:overlap}
\end{figure*}

\newcommand{\Lone}{3.6}
\newcommand{\Ltwo}{4.8}
\newcommand{\Lthree}{6.0}
\newcommand{\Lfour}{7.9}
\newcommand{\Lfive}{9.8}
\newcommand{\Nonegt}{957,095}
\newcommand{\Ntwogt}{718,542}
\newcommand{\Nthreegt}{478,585}
\newcommand{\Nfourgt}{239,567}
\newcommand{\Nfivegt}{119,353}

\begin{table*}
\centering
\begin{tabular}{|c|c|c|c|c|c|c|}
\cline{4-7}
\multicolumn{3}{l|}{} 
& \multicolumn{2}{c|}{f150 $\&$ ftot} 
& \multicolumn{1}{c|}{ILC}
& \multicolumn{1}{c|}{ILC (z $<$ 0.8)}
\\ \hline
Bin & Luminosity cut & Mass cut $M_{\mathrm{vir}}$ 
& \multirow{2}{*}{$N_{\mathrm{gal}}$} 
& \multirow{2}{*}{$\langle z\rangle$}
& \multirow{2}{*}{$N_{\mathrm{gal}}$}
& \multirow{2}{*}{$N_{\mathrm{gal}}$}
\\
Label & ($10^{10}L_{\odot}$) & ($10^{13}M_{\odot}$)
&  &  & &
\\ \hline
L36  & $L>\Lone$         & $M>0.38$        & \Nonegt   & 0.76 & 913,286 & 512,437 \\
L48  & $L>\Ltwo$         & $M>0.63$        & \Ntwogt   & 0.78 & 685,766 & 364,878 \\
L60  & $L>\Lthree$       & $M>0.96$        & \Nthreegt & 0.81 & 456,803 & 219,341 \\
L79  & $L>\Lfour$        & $M>1.65$        & \Nfourgt  & 0.85 & 228,650 & 92,729  \\
L98  & $L>\Lfive$        & $M>2.59$        & \Nfivegt  & 0.88 & 114,024 & 40,387  \\
L36D & $\Lone<L<\Ltwo$   & $0.38<M<0.63$   & 238,553   & 0.71 & 227,520 & 147,559 \\
L48D & $\Ltwo<L<\Lthree$ & $0.63<M<0.96$   & 239,957   & 0.73 & 228,963 & 145,537 \\
L60D & $\Lthree<L<\Lfour$& $0.96<M<1.65$   & 239,018   & 0.77 & 228,153 & 126,612 \\
$\quad$L79D$\quad$ 
& $\quad \Lfour<L<\Lfive \quad$ 
& $\quad 1.65<M<2.59 \quad$   
& $\quad$ 120,214 $\quad$ 
& $\quad$ 0.81 $\quad$ 
& $\quad$ 114,626 $\quad$ 
& $\quad$ 52,342 $\quad$ \\
\hline
\end{tabular}
\caption{Overview of the nine luminosity-selected samples used in this study (5 cumulative and 4 disjoint), along with the corresponding bin labels referenced throughout the paper. For each sample we list the host halo mass range ($M_{vir}$), the number of galaxies ($N_{\mathrm{gal}}$), and the average redshift ($\langle z \rangle$). The number of galaxies for both ILC and ILC ($z < 0.8$) samples are also listed.}
\label{tab1}
\end{table*}


The Dark Energy Spectroscopic Instrument (DESI) is conducting a spectroscopic survey designed to investigate the nature of dark energy by mapping the three-dimensional distribution of over 40 million galaxies and quasars \cite{DESI:2016fyo,DESI:2016igz,DESI:2022xcl} with precise redshift measurements, and is expected to reach a total of over 60 million targets. The resulting large-scale structure data will allow for high-precision studies of baryon acoustic oscillations and redshift-space distortions, providing critical constraints on cosmological models and the expansion history of the Universe \cite{DESI:2013agm, Ishak:2024jhs}. Following the DESI Survey Validation phase \cite{DESI:2023dwi}, the Early Data Release (EDR) \cite{DESI:2023ytc} and the First Data Release (DR1) \cite{DESI:2025fxa} were completed. 

We use the Luminous Red Galaxies (LRGs) sample  \cite{Zhou:2020mgr,DESI:2022gle,Zhou:2023gji} from DR1. Our spectroscopic LRG sample is selected following the DESI Key-Project (KP) \cite{DESI:2024uvr}. The selection provides 2,388,542 galaxies with luminosities derived from the absolute magnitudes listed in the DESI Value Added Catalog (VAC) \cite{2023ascl.soft08005M} that is constructed by jointly fitting DESI’s observed-frame optical spectra and broadband photometry with physically motivated stellar population synthesis and emission-line templates to derive galaxy physical properties. Within the sample, 1,689,748 galaxies overlap with the ACT DR6 f150 map. Following \cite{Calafut:2021wkx}, we performed an inverse white noise variance cut at a level of 45 $\mu K$ per pixel to the overlapping sample, which removes 69,406 galaxies. Galactic plane masking was then applied using the mask that is used in generating the 2015 \textit{Planck} Compton-y map \cite{Planck:2015vgm} to minimize contamination from the Milky Way. A conservative 50\% mask was chosen to exclude sources located near the Galactic plane, which led to the removal of 392,003 additional sources. We use a 50\% cut, consistent with the criterion adopted in previous work \cite{Calafut:2021wkx}. Using a larger cut such as 70\% would significantly reduce the number of objects for analysis. We finally mask point sources using the point source and cluster catalog used in the ACT DR6 map production process. Following \cite{Vavagiakis:2021ilq}, a 15 mJy point source mask at $10^\prime$ radius was used in the Deep 56 (D56) region \cite{2016SPIE.9910E..14D}, a 680 sq. deg. deep sky patch from the ACT surveys, while a 100 mJy mask at $5^\prime$ radius was applied in the rest of the areas. The point source mask removes 7,071 sources from the sample. Using the same point source and cluster catalog and following \cite{Liu:2025zqo}, we also mask out LRGs that are close to massive tSZ clusters as very massive clusters can dominate and skew the measurements. The final catalog selection, after applying all masks, contains 1,197,347 galaxies. The redshift distribution of this sample is shown in Fig.~\ref{fig:redshift} and in Fig.~\ref{fig:overlap} we present the overlap between our selected sample and the ACT DR6 f150 CMB map. In Fig. \ref{fig:redshift}, we also show the redshift distribution of the LRGs sample previously studied in \cite{Calafut:2021wkx} for comparison. The DESI DR1 sample, extending from $0<z<1.5$, with a peak around $z=0.8$, probes deeper into the Universe relative to SDSS DR15, which was confined to lower redshifts, covering $0<z<0.8$ and peaking at $z=0.55$. Most pairwise kSZ analyses to date (e.g., \cite{Hand_2012, DeBernardis:2016pdv, DES:2016umt, Chen:2021pwg, Calafut:2021wkx}) have focused on SDSS galaxy cluster samples at relatively low redshifts, $z<0.8$. With its extensive sky coverage, high target density, and precise redshift measurements, DESI enables the systematic identification and study of  tracer galaxies and galaxy clusters to significantly higher redshifts, extending well beyond $z>1$. In this work, in addition to studying the full dataset, we will also perform an analysis of the $z < 0.8$ subsample on the ILC map to ensure continuity with previous studies. 

We divide the sample by luminosity using the 20th, 40th, 60th, and 80th percentiles, corresponding to luminosities L$(10^{10} L_{\odot}) $= 3.6, 4.8, 6.0 and 7.9, referred to as L36, L48, L60, L79 respectively. We also consider the equivalent discrete bins and additionally include a sample with a luminosity cut of $9.8 \times 10^{10} L_{\odot}$, enabling a continuous comparison with the sample defined in \cite{Calafut:2021wkx}. The properties of the samples in each luminosity-selected bin are outlined in Table~\ref{tab1}, which includes luminosity-inferred mass ranges, the number of galaxies, and the average redshift and luminosities of each bin. We notice that the ILC map has a slightly different footprint from the f150 and ftot maps near the edges. Therefore, we exclude some objects in the ILC sample that lie close to the apodized map boundaries. The mass estimates are derived following the approach described in \cite{Vavagiakis:2021ilq}: stellar masses are first computed from galaxy luminosities using the mass-to-light ratio \cite{Bell:2000jt,Bell:2003cj,Chabrier:2003ki}, and halo masses are then inferred via the stellar–halo mass relation \cite{Kravtsov:2014sra}. Our primary analysis uses the L36 sample of 957,095 galaxies. Note, by comparison, that the previous analysis in \cite{Vavagiakis:2021ilq,Calafut:2021wkx} considered 343,647 galaxies with the lowest luminosity cut of $4.3 \times 10^{10} L_{\odot}$. 

\section{Methods}
\label{sec:methods}

\subsection{Secondary Anisotropies}
\label{sec:obs}

The kSZ effect arises from inverse Compton scattering of CMB photons off electrons moving with bulk velocities, imprinting anisotropies proportional to the line-of-sight momentum of the ionized gas, and is given by:
\begin{equation}
\label{eq:kSZ}
    \frac{\Delta T_{kSZ}}{T_{CMB}} = -\frac{\sigma_T}{c} \int n_e (v \cdot \hat{n}) dl,
\end{equation}
where $\sigma_T$ is the Thomson cross-section, $n_e$ is the electron number density, $v$ is the peculiar velocity of the electrons, and $\hat{n}$ is the line-of-sight unit vector. 

In contrast, the tSZ effect originates from inverse Compton scattering of CMB photons by hot electrons within galaxy clusters, altering the CMB's spectral energy distribution. The resulting temperature change is given by:
\begin{equation}
    \frac{\Delta T_{tSZ}}{T_{CMB}} = f(\nu) y,
\end{equation}
and
\begin{equation}
\label{eq:tSZ}
    y = \frac{\sigma_{T} k_B}{m_e c^2}\int n_e T_e dl,
\end{equation}
where $f(\nu) = x \coth(x/2) - 4$, $x = \frac{h\nu}{k_B T_0}$, $y$ is the Compton parameter, $T_e$ is the electron temperature, $m_e$ is the electron mass, and $k_B$ is Boltzmann's constant. 

Gravitational lensing of CMB photons by intervening structures distorts the observed anisotropies through lensing convergence $\kappa$, defined as: 
\begin{equation}
\label{eq:kappa}
    \kappa(\hat{n}) = \frac{3H_0^2 \Omega_m}{2c^2} \int_0^{\chi_*} d\chi\, \frac{\chi (\chi_* - \chi)}{\chi_*} \frac{\delta(\chi \hat{n}, z)}{a(\chi)},
\end{equation}
where $H_0$ is the Hubble constant, $\omega_m$ is the matter density parameter, $\chi$ is the comoving radial distance, $\chi_*$ is the comoving radial distance to the last scattering surface, $\delta$ is the density contrast, and $a(\chi)$ is the scale factor. This formalism is applied in Sec. \ref{sec:Phat}.

We measure the kSZ and tSZ signal using aperture photometry (AP). The method involves computing the temperature decrement in a central aperture and subtracting the average temperature in a surrounding annulus of equal area. The aperture photometry estimate of the signal at the location of a galaxy cluster is given by:
\begin{equation}
\label{eq:AP}
    T_{AP} = \frac{1}{N_{\text{ap}}} \sum_{i \in \text{ap}} T_i - \frac{1}{N_{\text{ann}}} \sum_{j \in \text{ann}} T_j,
\end{equation}
where \( T_i \) are the temperature values of the pixels within the aperture, \( T_j \) are the temperature values of the pixels in the annulus, and \( N_{\text{ap}} \) and \( N_{\text{ann}} \) are the number of pixels in the aperture and annulus, respectively.

For the kSZ temperature, following \cite{Hand_2012, DeBernardis:2016pdv, Calafut:2021wkx}, we use an aperture size equal to 2.1$^\prime$,which is also a good fit for our sample. We have also tested alternative aperture sizes of $1.5^\prime$ and $2.5^\prime$ around the fiducial value of $2.1^\prime$, and find that the signal-to-noise ratio peaks at $2.1^\prime$, consistent with previous findings \cite{Flender:2015btu,Gong:2023hse}. We also subtract a redshift smoothed temperature, $\bar{T}_{AP}$, to the AP measurement $T_{AP}$. The aim is to eliminate possible contamination that could imitate a pairwise signal when comparing aperture temperatures of objects at different redshifts, and it is given by:
\begin{equation}
\label{eq:APz}
    \bar{T}_{AP}(z_i) = 
\frac{\sum_j T_{AP}(z_i) \exp\left( -\frac{(z_i - z_j)^2}{2\sigma_z^2} \right)}
     {\sum_j \exp\left( -\frac{(z_i - z_j)^2}{2\sigma_z^2} \right)},
\end{equation}
where we use $\sigma_z = 0.01$ as in \cite{DeBernardis:2016pdv, Calafut:2021wkx}.

As discussed in \cite{Gong:2023hse}, aperture photometry can suffer systematic signal attenuation, particularly when signals of interest extend into the background annulus. Since the annulus is typically used to estimate and subtract the local background, any genuine signal present in this region is mistakenly removed, leading to an underestimation of the true source amplitude. Following the method described in \cite{Gong:2023hse}, we also calculate the AP correction factor, $A_{AP}$, to account for the signal attenuation introduced by background subtraction. Applying $A_{AP}$, we can recover an unbiased measurement of the kSZ signal from the central disk , mitigating the biases that arise from the overlap of extended kSZ signal profiles with the annulus.

\subsection{Pairwise statistic}
\label{sec:pariwise}
The pairwise velocity statistic is a fundamental tool for probing the large-scale dynamics of cosmic structure. Defined as the mean relative velocity between pairs of galaxies (clusters) as a function of their comoving separation, the pairwise velocity is given by:
\begin{equation}
\label{eq:Vhat}
    \hat{V}(r) = -\frac{\sum_{i<j} (v_i - v_j) c_{ij}}{\sum_{i<j} c_{ij}^2},
\end{equation}
where $r$ is the comoving separation between two objects and $v$ is the line-of-sight velocity of a cluster. The geometric weight $c_{ij}$ projects the velocity difference along the line of sight and is formulated as follows:
\begin{equation}
    c_{ij} = \hat{\mathbf{r}}_{ij} \cdot \frac{\hat{\mathbf{r}}_i + \hat{\mathbf{r}}_j}{2} = \frac{(r_i - r_j)(1 + \cos\alpha)}{2\sqrt{r_i^2 + r_j^2 - 2 r_i r_j \cos\alpha}},
\end{equation}
where $\hat{\mathbf{r}}_i$ and $\hat{\mathbf{r}}_j$ are unit vectors in the direction of objects $i$ and $j$, $r_i$ and $r_j$ are their comoving distances, and $\cos\alpha = \hat{\mathbf{r}}_i \cdot \hat{\mathbf{r}}_j$ defines the angle between the two lines of sight. The vector $\hat{\mathbf{r}}_{ij}$ denotes the unit separation vector between the pair.

In practice, this velocity signal is often inferred from the kSZ effect. The pairwise kSZ momentum estimator is given by:
\begin{equation}
\label{eq:Phat}
    \hat{p}_{\mathrm{kSZ}}(r, z_i) = -\frac{\sum_{i<j} \left(\Delta T_{\mathrm{kSZ}, i} - \Delta T_{\mathrm{kSZ}, j} \right) c_{ij}}{\sum_{i<j} c_{ij}^2},
\end{equation}
where $\Delta T_{\mathrm{kSZ}, i}$ is the observed temperature fluctuation at object $i$ due to the kSZ effect, computed as $T_{AP} - \bar{T}_{AP}$. The sums in Eqs. \ref{eq:Phat}, \ref{eq:APz}, \ref{eq:AP} are implemented in the pipeline described in Ref. \cite{Gallardo:2025lzu}. 

\begin{figure*}
    \centering
    \includegraphics[width=0.9\textwidth]{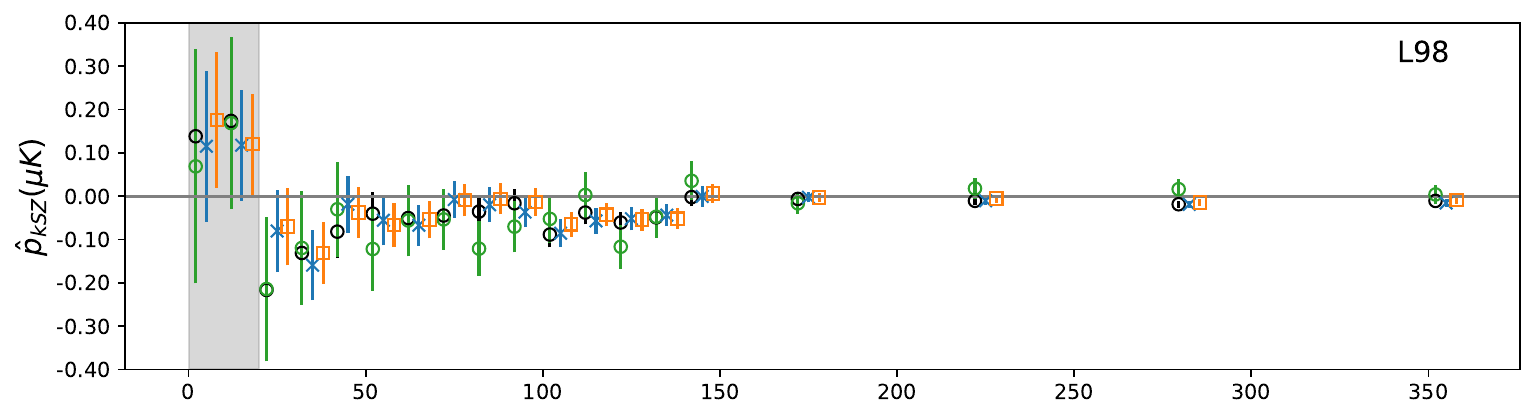}
    \includegraphics[width=0.9\textwidth]{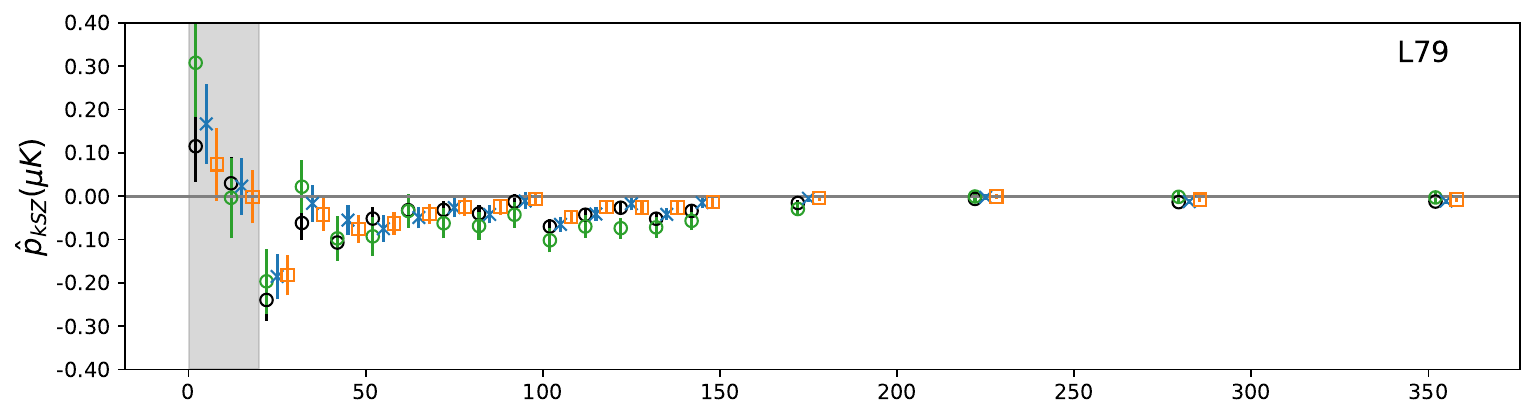}
    \includegraphics[width=0.9\textwidth]{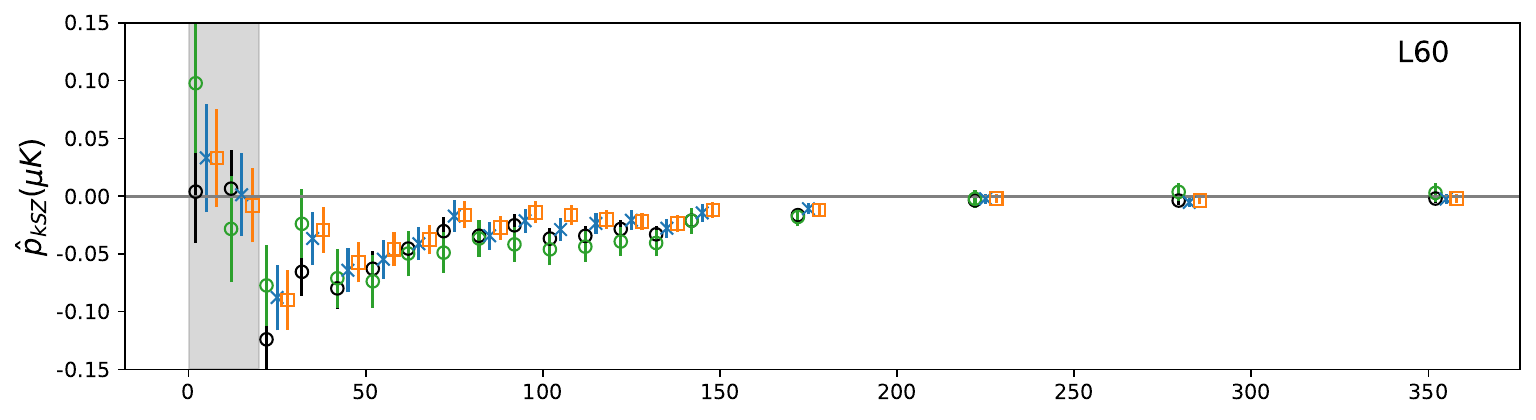}
    \includegraphics[width=0.9\textwidth]{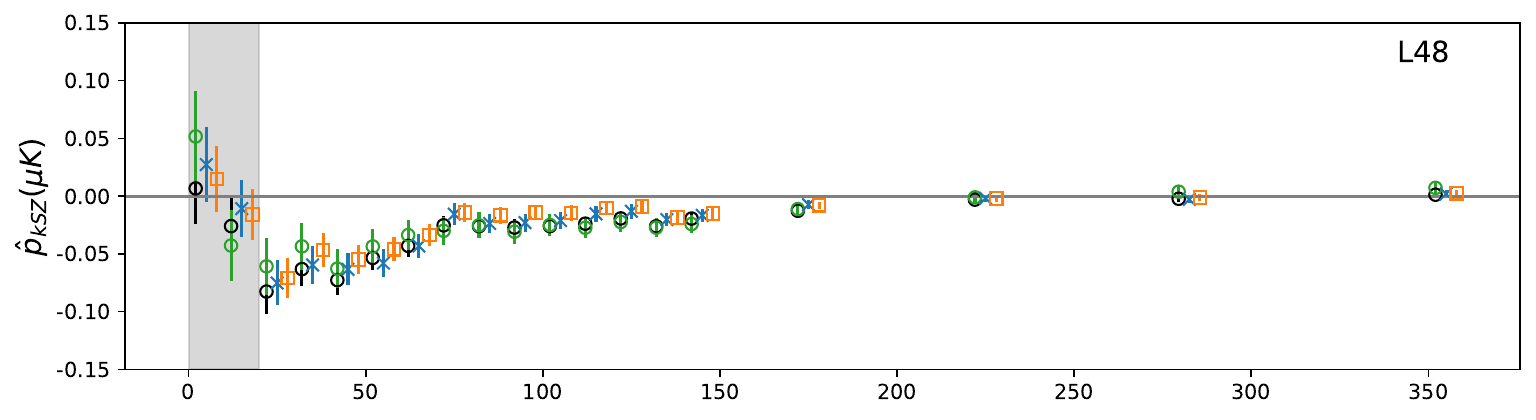}
    \includegraphics[width=0.9\textwidth]{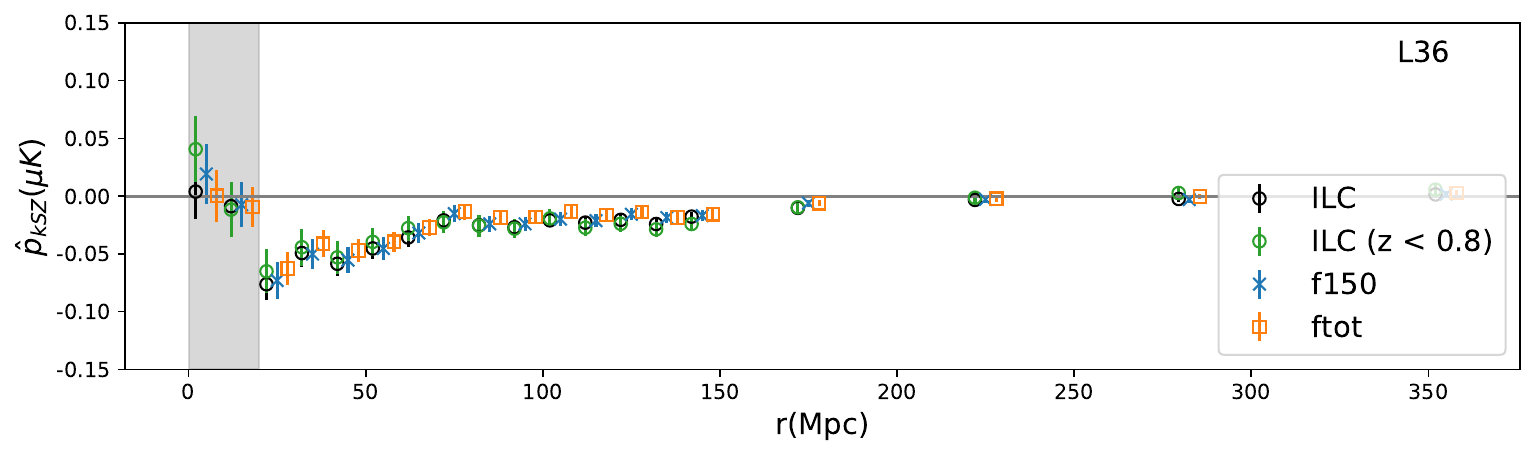}
    \caption{Measured pairwise kSZ momentum curves for the ILC [black circle], $z < 0.8$ ILC subsample [green], f150 [blue cross], and ftot [orange square] maps, based on sources in nine luminosity-selected galaxy tracer samples. From top to bottom, these samples are: L36, L48, L60, L79, and L98. Two different vertical scales are used, scaled to their respective amplitudes. $1\sigma$ uncertainty error bars obtained through bootstrap analysis are also presented. In this study, we concentrate on scales $r > 20$ Mpc, and consequently shade the region corresponding to $r < 20$ Mpc.}
\label{fig:Phat_result}
\end{figure*}

\begin{figure*}
    \centering
    \includegraphics[width=0.9\textwidth]{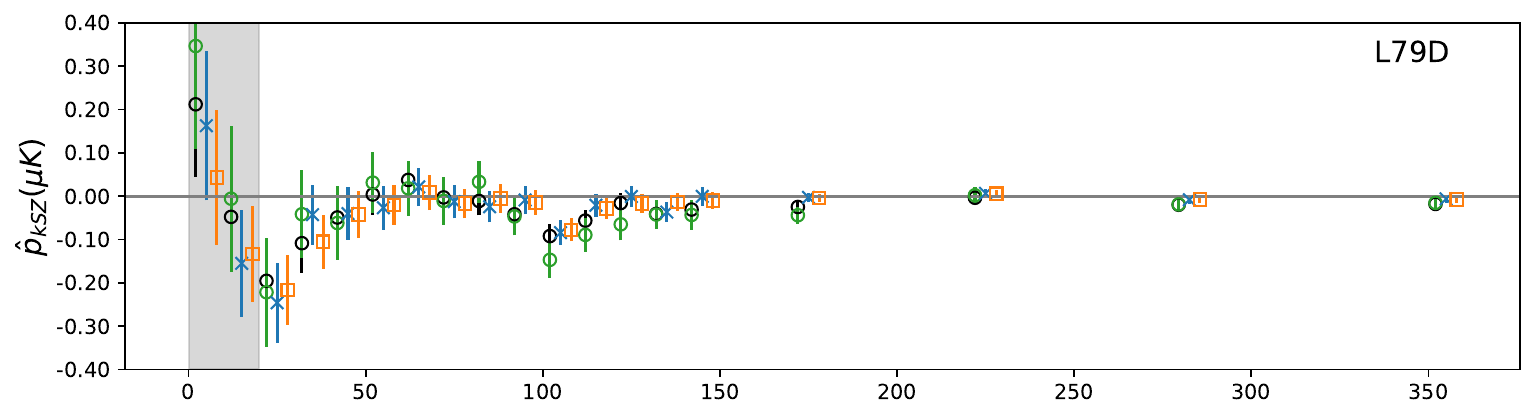}
    \includegraphics[width=0.9\textwidth]{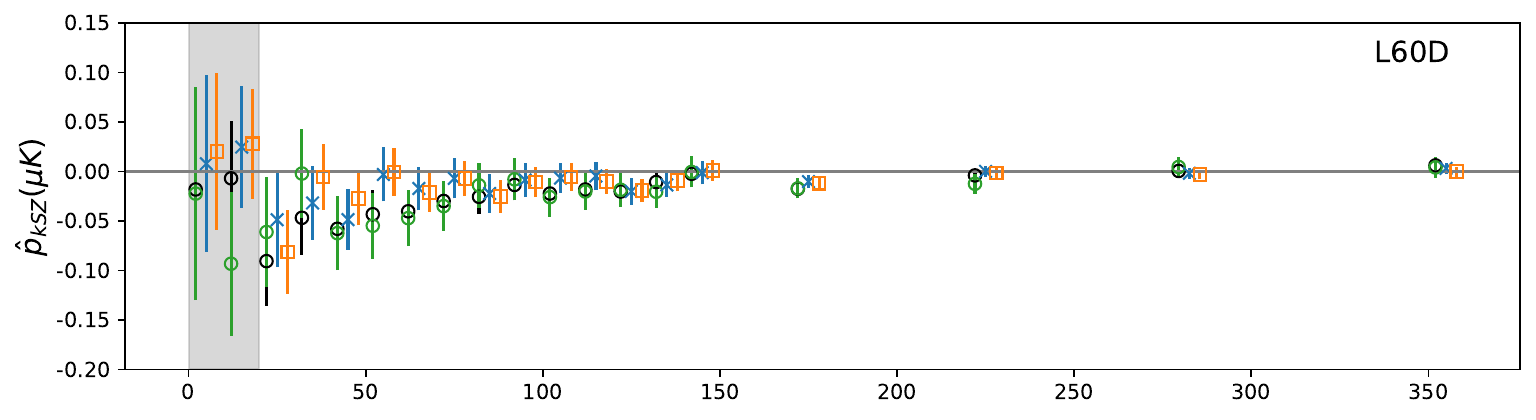}
    \includegraphics[width=0.9\textwidth]{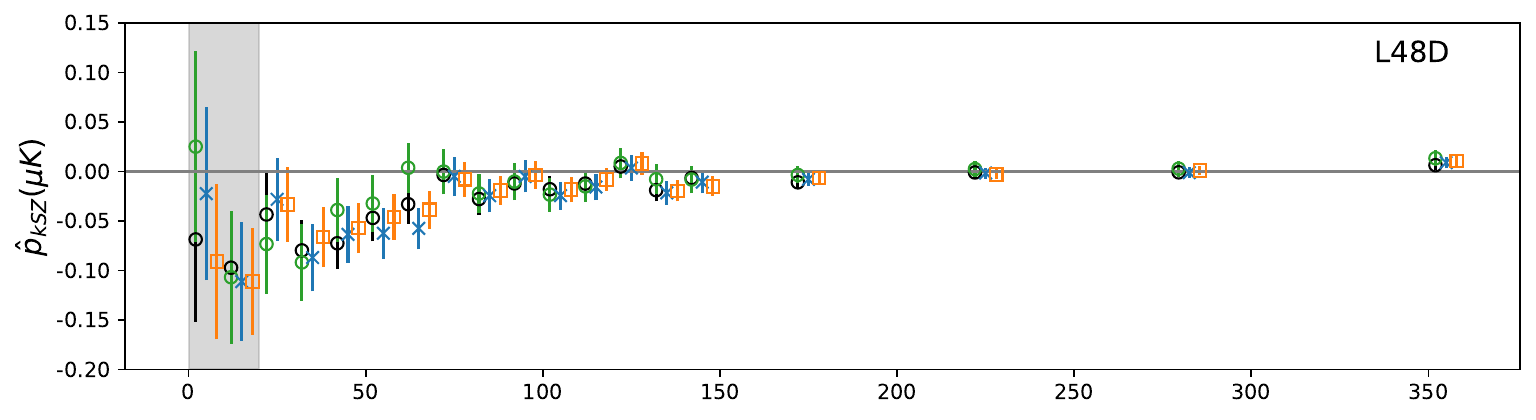}
    \includegraphics[width=0.9\textwidth]{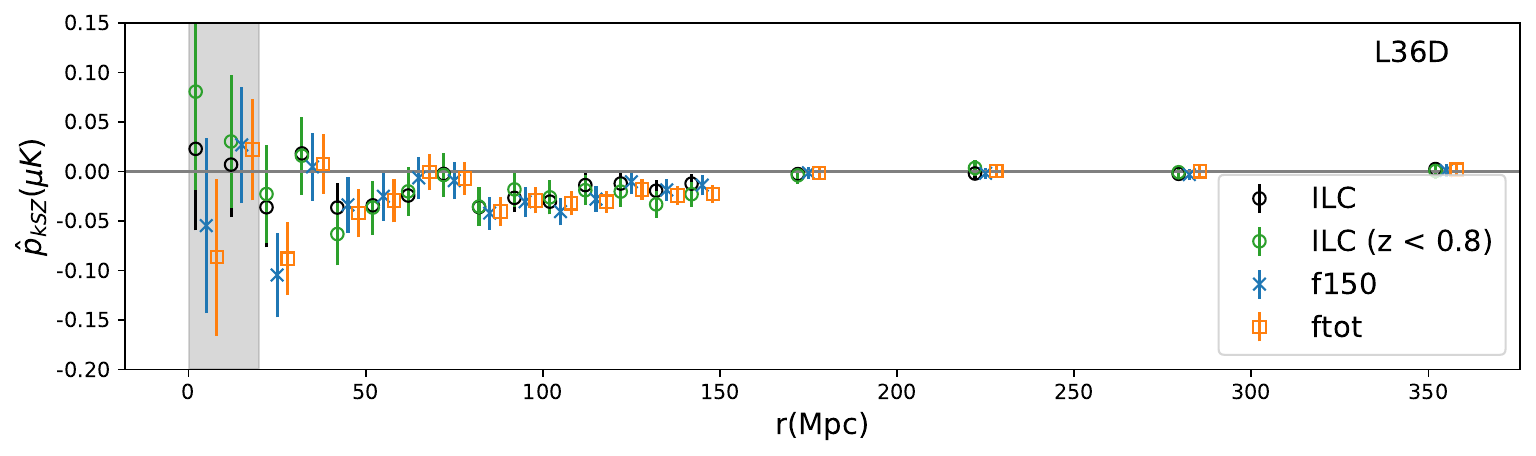}
    \caption{Same as Fig. \ref{fig:Phat_result}, but for disjoint luminosity samples: L36D, L48D, L60D, and L79D. Two different vertical scales are used, scaled to their respective amplitudes.} 
    \label{fig:Phat_d_result}
\end{figure*}

In accordance with \cite{Calafut:2021wkx}, our analysis is conducted using radial separation bins (e.g. used in Fig.\ref{fig:Phat_result} and \ref{fig:Phat_d_result}), starting with evenly spaced intervals of 10 Mpc centered between 5 and 145 Mpc. In addition, we include four broader, non-uniform bins centered at 175, 225, 282.5, and 355 Mpc.

We also model a theoretical prediction for the observed pairwise velocity using linear theory \cite{Sheth:2000ff} and the best-fit \textit{Planck} cosmological
parameters \cite{Planck:2018vyg}. Following \cite{Mueller:2014dba,Mueller:2014nsa}, it is calculated as 
\begin{equation}
    \label{eq:Vhat_theo}
    V(r, z) = -\frac{2}{3} \frac{f(z) H(z) r}{1 + z} \frac{\bar{\xi}_h(r, z)}{1 + \xi_h(r, z)},
\end{equation}
where \( f(z) \) denotes the linear growth rate and \( H(z) \) is the Hubble parameter. The functions \( \xi_h \) and \( \bar{\xi}_h \) represent the 2-point halo correlation function and its volume-averaged form, respectively \cite{Sheth:2000ff,Bhattacharya:2007sk}:
\begin{equation}
\xi_h(r, z) = \frac{1}{2\pi^2} \int dkk^2 j_0(kr) P(k, z) b_h^{(2)}(k),
\end{equation}

\begin{equation}
\bar{\xi}_h(r, z) = \frac{3}{r^3} \int_0^r dr' r'^2 \xi(r', z) b_h^{(1)}(k).
\end{equation}

Here, \( P(k, z) \) is the linear matter power spectrum and \( j_0(x) = \sin(x)/x \) is the zeroth-order spherical Bessel function. The quantity \( b_h^{(q)} \) denotes the \( q \)-th order mass-averaged halo bias moment, defined as:

\begin{equation}
b_h^{(q)}(z) = \frac{\int_{M_{\text{min}}}^{M_{\text{max}}} dM\, M\, n(M, z)\, b^q(M)\, W^2[kR(M, z)]}{\int_{M_{\text{min}}}^{M_{\text{max}}} dM\, M\, n(M, z)\, W^2[kR(M, z)]},
\end{equation}
where \( n(M, z) \) is the number density of halos of mass \( M \), based on the halo mass function from \cite{Bhattacharya_2011}. The top-hat window function is:
\[
W(x) = 3 \frac{\sin x - x \cos x}{x^3}.
\]
The characteristic scale of a halo with mass \( M \) is:
\[
R(M, z) = \left[ \frac{3M}{4\pi \bar{\rho}(z)} \right]^{1/3},
\]
with \( \bar{\rho}(z) \) being the average matter density of the Universe. This scale represents the physical size of a dark matter halo of mass $M$ at a given redshift $z$.

\subsection{Mass-averaged optical depth estimate}
\label{sec:tau}
The theoretical expression for the observed pairwise kSZ momentum is given by
\begin{equation}
    \label{eq:P_V}
    \hat{p}_{\mathrm{th}}(r, z) = -\frac{T_{\mathrm{CMB}}}{c} \, \bar{\tau} \, \hat{V}(r, z),
\end{equation}
where $\bar{\tau}$ denotes the effective mass-averaged optical depth across the cluster samples and $z$ is the mean redshift. We compare the theoretical predictions for the pairwise velocity with our measurements of the pairwise kSZ momentum. Since the derived  $\bar{\tau}$ values are computed by dividing by the theoretical pairwise velocity, they are contingent on the accuracy of the theoretical pairwise velocity values. We omit two bins with $r<20$ Mpc as nonlinear velocity effects, not accounted for in the linear theoretical model for $\hat{V}$ described in (\ref{eq:Vhat_theo}), become significant at these scales. We have performed a robustness test varying the minimum separation used in the fit, $r_{\min}=20,30,40\,\mathrm{Mpc}$, and find that the best-fit $\bar{\tau}_{\mathrm{AP}}$ shifts by $<3\%$ across these choices, consistent with the expected (few--$10\%$) level of non-linear corrections \cite{Reid_2011,DES:2016umt,DeBernardis:2016pdv} in this separation range and well within our statistical uncertainties.

We use the chi-squared statistic to evaluate the likelihood of effective mass-weighted optical depth $\bar{\tau}$,
\begin{equation}
    \chi^2(\bar{\tau}) = \sum_{ij} \Delta \hat{p}_i(\bar{\tau}) \hat{C}^{-1}_{ij} \Delta \hat{p}_j(\bar{\tau}),
\end{equation}
where \( \Delta \hat{p}_i(\bar{\tau}) = \hat{p}_{i,\text{th}}(\bar{\tau}) - \hat{p}_{i,\text{obs}} \), with \( \hat{p}_{i,\text{th}}(\bar{\tau}) \) representing the theoretical kSZ pairwise momentum prediction at cluster separation \( r_i \) for a given average optical depth \( \bar{\tau} \), and \( \hat{p}_{i,\text{obs}} \) being the observed value. $\hat{C}^{-1}_{ij}$ is the inverse of the covariance.
A comparison of jackknife, bootstrap and simulation-derived covariance estimation methods for the ACT-SDSS pairwise statistic \cite{Calafut:2021wkx} found that the covariance matrix estimated by bootstrap analysis with 1000 iterations was in good agreement with that estimated from noise simulations while the jackknife resampling overestimates the covariance (possibly because the resampling by excision of subsamples rather than replacement can remove one of the clusters in pairs that span across subsamples leading to a biased estimate). We therefore use the same bootstrap method to estimate the covariance, $\hat{C}_{ij}$, here.  Under the null hypothesis of no kSZ signal, we have $\bar{\tau} = 0$. While we report SNR evaluated relative to the zero-signal case for comparison, throughout the paper we quote the detection significance evaluated at the best-fit signal amplitude as the headline detection significance.

The best-fit value is obtained at \(\chi^2_{\text{min}} \), and we compute the Probability-To-Exceed (PTE), which quantifies the probability of obtaining a larger chi-squared value:
\begin{equation}
    \text{PTE} = \int_{\chi^2_{\text{min}}}^{\infty} \chi^2_m(x) \, dx,
\end{equation}
where \( \chi^2_m \) is the chi-squared distribution with \( m \) degrees of freedom. 

In accordance with \cite{Calafut:2021wkx}, the signal-to-noise ratio (SNR) is calculated assuming the signal is given by the best-fit theoretical model:
\begin{equation}
\text{SNR}(\bar{\tau}) = \sqrt{ \sum_{ij} \hat{p}_{i,\text{th}}(\bar{\tau}) \hat{C}^{-1}_{ij} \hat{p}_{j,\text{th}}(\bar{\tau}) }.
\end{equation}

It is important to note that we use aperture photometry to measure the kSZ temperature. Consequently, the pairwise kSZ curves calculated from $T_{AP}$ will also be attenuated, as discussed in Sec.~\ref{sec:obs}. Therefore, the best-fit optical depth derived using the method described in this section will also be attenuated and is denoted as $\bar{\tau}_{AP}$.

\begin{table*}
\centering
\resizebox{0.98\textwidth}{!}{%
\begin{tabular}{|c|c|c|c|c|c|c|c|c|c|c|c|c|c|c|c|c|}
\hline
\multirow{2}{*}{\begin{tabular}[c]{@{}c@{}}Tracer\\ sample\end{tabular}} & 
\multicolumn{4}{c|}{ftot} & 
\multicolumn{4}{c|}{f150} & 
\multicolumn{4}{c|}{ILC} & 
\multicolumn{4}{c|}{ILC ($z < 0.8$)} \\
\cline{2-17}
 & $\bar{\tau}_{AP}$ ($\times 10^{-4}$) & $\chi^2_{\text{min}}$ & PTE & SNR & 
   $\bar{\tau}_{AP}$ ($\times 10^{-4}$) & $\chi^2_{\text{min}}$ & PTE & SNR & 
   $\bar{\tau}_{AP}$ ($\times 10^{-4}$) & $\chi^2_{\text{min}}$ & PTE & SNR & 
   $\bar{\tau}_{AP}$ ($\times 10^{-4}$) & $\chi^2_{\text{min}}$ & PTE & SNR \\
\hline

L36 & 
0.36 $\pm$ 0.05 & 18 & 0.33 & 7.2 & 
0.42 $\pm$ 0.06 & 18 & 0.29 & 7.5 & 
0.46 $\pm$ 0.05 & 18 & 0.33 & 9.3 & 
0.41 $\pm$ 0.07 & 19 & 0.26 & 5.9 \\

L48 & 
0.37 $\pm$ 0.06 & 15 & 0.51 & 6.8 & 
0.43 $\pm$ 0.06 & 16 & 0.45 & 7.1 & 
0.51 $\pm$ 0.06 & 13 & 0.70 & 9.2 & 
0.43 $\pm$ 0.08 & 12 & 0.75 & 5.3 \\

L60 & 
0.39 $\pm$ 0.07 & 15 & 0.50 & 5.7 & 
0.44 $\pm$ 0.08 & 15 & 0.53 & 5.3 & 
0.54 $\pm$ 0.07 & 16 & 0.47 & 8.1 & 
0.54 $\pm$ 0.11 & 15 & 0.54 & 4.9 \\

L79 & 
0.43 $\pm$ 0.10 & 20 & 0.22 & 4.2 & 
0.49 $\pm$ 0.11 & 29 & 0.05 & 4.1 & 
0.58 $\pm$ 0.10 & 25 & 0.05 & 5.4 & 
0.72 $\pm$ 0.19 & 23 & 0.14 & 3.9 \\

L98 & 
0.45 $\pm$ 0.16 & 15 & 0.55 & 2.8 & 
0.55 $\pm$ 0.18 & 15 & 0.50 & 3.1 & 
0.62 $\pm$ 0.16 & 15 & 0.56 & 3.9 & 
0.72 $\pm$ 0.32 & 13 & 0.70 & 2.3 \\ \hline

L36D & 
0.36 $\pm$ 0.10 & 21 & 0.17 & 3.6 & 
0.36 $\pm$ 0.12 & 16 & 0.45 & 3.1 & 
0.27 $\pm$ 0.11 & 12 & 0.74 & 2.5 & 
0.28 $\pm$ 0.13 & 13 & 0.68 & 2.1 \\

L48D & 
0.36 $\pm$ 0.10 & 18 & 0.35 & 3.7 & 
0.43 $\pm$ 0.11 & 17 & 0.40 & 4.0 & 
0.41 $\pm$ 0.10 & 13 & 0.68 & 4.1 & 
0.35 $\pm$ 0.13 & 12 & 0.76 & 2.6 \\

L60D & 
0.23 $\pm$ 0.10 & 10 & 0.88 & 2.3 & 
0.25 $\pm$ 0.11 & 7 & 0.98 & 2.3 & 
0.45 $\pm$ 0.11 & 10 & 0.90 & 4.2 & 
0.38 $\pm$ 0.14 & 10 & 0.91 & 2.7 \\

L79D & 
0.36 $\pm$ 0.16 & 17 & 0.39 & 2.2 & 
0.34 $\pm$ 0.18 & 19 & 0.27 & 1.8 & 
0.51 $\pm$ 0.17 & 25 & 0.07 & 3.2 & 
0.54 $\pm$ 0.26 & 24 & 0.10 & 2.2 \\ \hline

\end{tabular}
}
\caption{Summary of the best-fit $\bar{\tau}_{AP}$ values and associated $1\sigma$ uncertainties for the DR6 ftot [left], f150 [center], ILC and ILC ($z<0.8$) maps [right], based on nine luminosity-selected galaxy tracer samples. The fitted values are obtained by comparing the observed pairwise kSZ curves with the theoretical pairwise velocity derived from the best-fit \textit{Planck} cosmology. The uncertainties are estimated using the bootstrap method. The corresponding $\chi^2$ values (for 17 degrees of freedom), signal-to-noise ratio (SNR), and probability to exceed (PTE) are also listed for each case.}
\label{tab2}
\end{table*}

\subsection{Peculiar velocity estimation}
\label{sec:v}

In addition to the effective mass-averaged optical depth, we also estimate the kSZ-weighted optical depth for each individual galaxy cluster using the machine learning method developed in \cite{Gong:2025gmc} (for more details, please refer to the paper). The machine learning model uses a Gradient Boosted Decision Trees (GBDT) algorithm to make prediction. GBDT is an ensemble learning method that builds a sequence of decision trees, where each new tree corrects the errors of the previous trees, improving accuracy by iteratively minimizing errors using gradient-based methods. The model is trained on the Simons Observatory Forecast simulation \cite{SimonsObservatory:2018koc}, which delivers a comprehensive suite of simulated kSZ, tSZ, and lensing convergence maps, together with an associated halo catalog. The method, as described in \cite{Gong:2025gmc}, uses a combination of model features measured from tSZ, using aperture photometry (in contrast to analyses in which the cluster tSZ signals are stacked), the lensing convergence, extracted via a machine-learning–based filtering model, and cluster mass achieved from the catalog to estimate the optical depth for the cluster independently.

This kSZ-weighted optical depth estimate is then used to combine with the kSZ temperature to estimate the peculiar velocity of each individual galaxy via
\begin{equation}
    \label{eq:vpred}
    v_{\text{pred}} = - \frac{c}{\tau_{\text{pred}}} \cdot \frac{{T}_{AP}}{T_{CMB}}.
\end{equation} 
Then, we use these derived velocities and (\ref{eq:Vhat}) to reconstruct the pairwise velocity statistic. Similar to the pairwise kSZ curves discussed above, the velocity is derived from $T_{AP}$. Consequently, the corresponding pairwise velocity is attenuated by a factor of $A_{AP}$ as discussed in Sec. \ref{sec:obs}.

\section{Results}
\label{sec:results}

\begin{figure*}
    \centering
    \includegraphics[width=\textwidth]{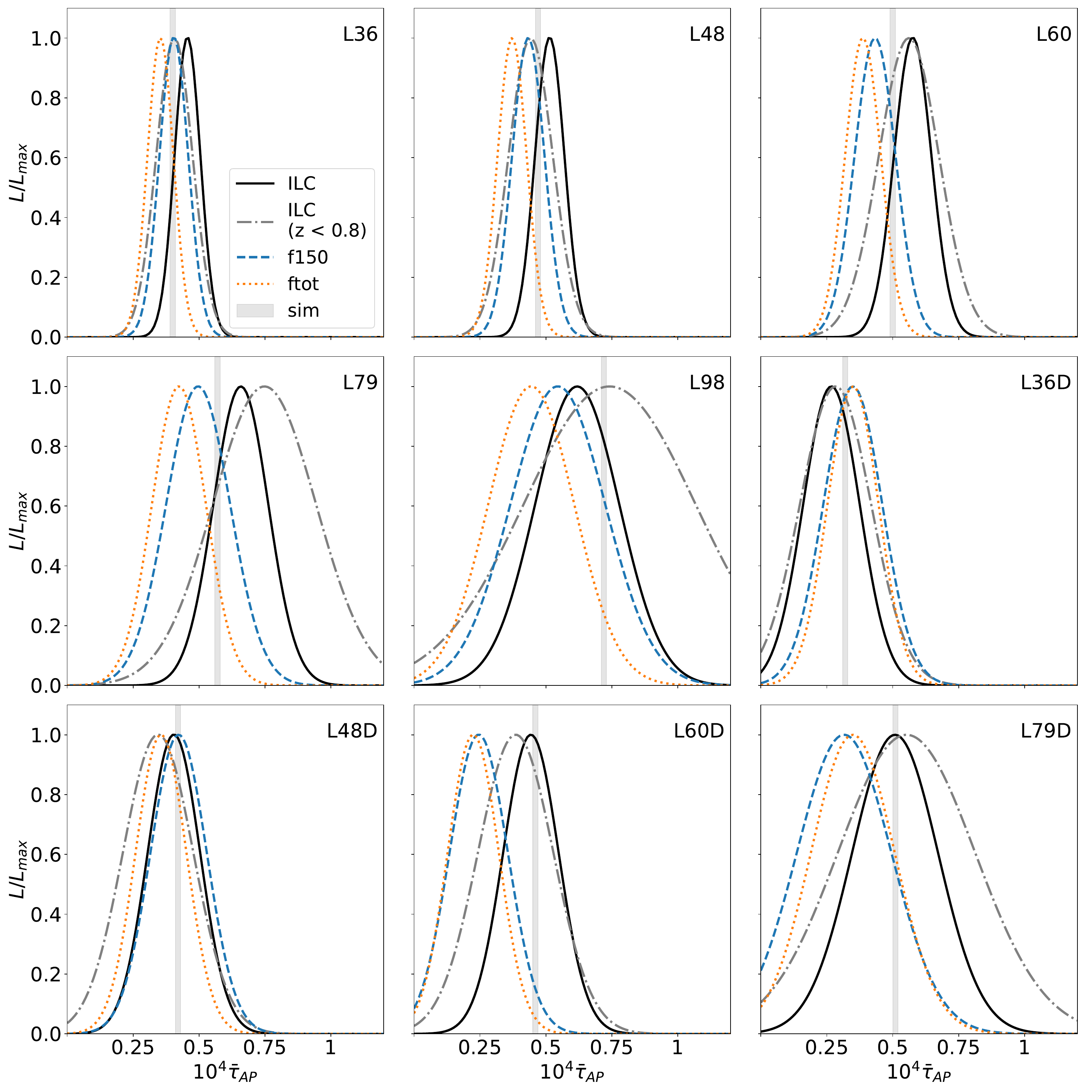}
    \caption{The normalized likelihoods, $L\propto \exp(-\chi^2)$, for the fitted values of $\bar{\tau}$ using the DR6 ILC [black solid] and its corresponding low redshift subsample [grey dashed], f150 [blue dashed], and ftot [orange dotted] maps, computed for each of the nine luminosity-based tracer samples. The optical depth derived from the simulations is also presented for reference.}
    \label{fig:likelihood}
\end{figure*}

\subsection{Pairwise kSZ momentum and mass-averaged optical
depth measurements}
\label{sec:Phat_tau}
In Fig.~\ref{fig:Phat_result} and \ref{fig:Phat_d_result}, we present the pairwise kSZ momentum curves derived from the ftot, f150, and ILC maps with 1$\sigma$ uncertainties derived from bootstrap analyses. For the L36, L48, L60, L79, and L98 samples, we find that the peak amplitude of the pairwise kSZ curves increases as the average luminosity of the sample increases, consistent with the expectation that the clusters formed in more massive halos possess greater optical depths and larger gravitational potentials and therefore larger pairwise velocities. On scales $r>20$Mpc, velocity correlations are theoretically predicted to be linear \cite{Vlah_2012, Hand_2012,Okumura:2013zva,Sugiyama:2015dsa,DeBernardis:2016pdv,Calafut:2021wkx}, and the negative pairwise signal is interpreted as evidence of gravitational infall between pairs of galaxy clusters as they are drawn toward one another due to their mutual gravitational attraction. As the physical separation between the cluster pairs increases, the strength of the gravitational interaction decreases, resulting in a weaker infall signal. For the discrete luminosity samples L36D, L48D, L60D, and L79D, we find that the kSZ signal becomes less significant as both average mass and sample size decrease.

For each luminosity-selected sample, the corresponding pairwise kSZ curves derived from the three  complementary maps (i.e., ftot, f150, and ILC) show agreement within the 1$\sigma$ uncertainty. This consistency suggests that the results are resistant to substantial and unique sources of contamination arising from foreground emissions that vary with frequency. In other words, even though each map employs observations in different effective frequencies, the agreement among the kSZ signal measurements implies that no single map is disproportionately affected by foreground interference, thereby affirming the reliability of the findings. These findings also extend to the low redshift sample (i.e., $z < 0.8$), and we also observe a good agreement with the full sample results within $1\sigma$ uncertainty, with no significant difference between the full and low redshift samples.

The pairwise kSZ signal is detected with high significance using the ILC map, with the SNR relative to the null signal varying based on the galaxy sample selection. We find that the detection significance improves as the sample size increases. The SNR increases from 5.6$\sigma$ for the L98 sample to 8.4$\sigma$ for L79, 8.9$\sigma$ for L60, and 9.9$\sigma$ for L48, ultimately peaking at 10.2$\sigma$ for the L36 sample. In the meantime, at roughly fixed sample size, as the sample luminosity increases the SNR for the corresponding discrete sample also increases ranging from 4.5$\sigma$ for L36D, up to 5.1$\sigma$ for L48D, 5.3$\sigma$ for L60D, and 6.0$\sigma$ for L79D.

Using Eq. (\ref{eq:P_V}), we estimate the effective mass-averaged optical depth, $\bar{\tau}_{AP}$, by comparing the measured pairwise kSZ momentum curves for $r>20$Mpc to the theoretical linear prediction given in (\ref{eq:Vhat_theo}). For each of the pairwise kSZ curves shown above, we summarize the best-fit mass-averaged optical depth constraints in Table~\ref{tab2}. The ILC map consistently gives the best detection of the mass-average optical depth over different luminosity bins compared with the other two maps. We find that the ILC map yields systematically higher $\bar{\tau}_{
AP}$ than f150/ftot across most luminosity bins. This is expected as the ILC map has a substantially different beam size, which can lead to consistently larger values of $\bar{\tau}_{AP}$, as discussed in \citep{Gong:2023hse}. The best measured detection of mass-averaged optical depth is obtained with the ILC map L36 sample, yielding SNR = 9.3 with $\chi^2 = 18$ for 17 degrees of freedom. This measurement achieves a 72$\%$ improvement in SNR compared with our previous work \cite{Calafut:2021wkx}, which used the ACT DR5 150 GHz CMB map and the SDSS DR15 LRG sample. The improvement is primarily driven by the increased sample size obtained from using the LRG sample from the DESI spectroscopic galaxy survey. The L48 sample achieves a comparable, but slightly lower, SNR of 9.2, but with a better $\chi^2 = 13$. The SNR increases as signal uncertainties decrease, primarily driven by the increasing number of galaxies in the luminosity bin.  The results demonstrate that the signal is not reliant on the smallest mass halo bin, but that bin does modestly improve the results through reducing statistical errors by increasing the sample size while contributing a smaller amplitude to the pairwise momentum signal. 

In Fig.~\ref{fig:likelihood}, we present the likelihood distributions of the fitted mass-averaged optical depth for each of the pairwise kSZ curves. The best-fit values are consistent across all three maps, indicating robustness in the measurement, which disfavors a dominant contribution from residual CIB or radio contamination (which would generically induce frequency-dependent shifts). In addition, \citep{Calafut:2021wkx} directly tested residual tSZ leakage by repeating the pairwise-kSZ/optical-depth measurement using an ILC map with tSZ deprojected and found no evidence for a bias relative to the baseline ILC case (with increased noise), providing an explicit constraint on potential tSZ contamination. As the sample size increases, the constraints on the optical depth become progressively tighter as progressively lower luminosity thresholds are applied. We also find that in most cases the best-fit mass-averaged optical depths are consistent between the low-redshift sample ($z < 0.8$) and the full sample, indicating that the inclusion of high redshift objects does not affect the measurements significantly. We find no statistically significant evidence for redshift evolution of optical depth within the redshift range probed by our sample. Similar observation is also reported in \cite{DES:2016umt} and the authors discussed that a
weak optical depth redshift evolution should be expected. This behavior is also qualitatively consistent with results from cosmological hydrodynamical simulations\cite{Flender:2015btu,SimonsObservatory:2018koc,Stein:2020its}, which generally predict only weak redshift evolution of the halo optical depth for samples selected at fixed mass or luminosity over similar redshift ranges. This is consistent with theoretical expectations for a fixed tracer selection, for which the average optical depth is primarily determined by the typical halo mass and baryon content of the sample. Since these quantities are not expected to evolve strongly over the redshift range considered, any intrinsic redshift evolution of the optical depth is expected to be weak and below our current sensitivity. In the figure, we also show the optical depth for each cluster sample, derived from a simulation \cite{Flender:2015btu}. We use simulated kSZ maps based on post-processed N-body simulations. The underlying N-body simulation adopts a WMAP7 cosmology and resolves halos identified with a friends-of-friends algorithm. For each halo, a single mass-weighted peculiar velocity is assigned and used to model the kSZ signal. The intracluster gas distribution follows a physically motivated gas prescription assuming hydrostatic equilibrium, and includes contributions from both halo gas and diffuse gas outside halos. This simulation-based optical depth corresponds to the best-fit value obtained by comparing the pairwise kSZ curves with the pairwise velocities measured from the simulated halo catalog and the simulated kSZ map. We are also using aperture photometry to measure the kSZ temperature from the simulated kSZ maps, and thus, the derived optical depth is $\bar{\tau}_{AP}$. We find that the simulation-derived optical depth agrees well with the ILC results, is slightly larger than the 150 GHz results, but remains well within the 1$\sigma$ uncertainty. This value serves only as a reference and as a cross-check of the robustness of our ACT $\times$ DESI pairwise kSZ–derived optical depth. A more detailed study of the ACT $\times$ DESI tSZ-derived optical depth will be presented in a forthcoming companion paper. In Appendix \ref{sec:appendix_a}, we also conduct a null test to support the significance of our measurements.

\begin{figure}
    \includegraphics[width=\columnwidth]{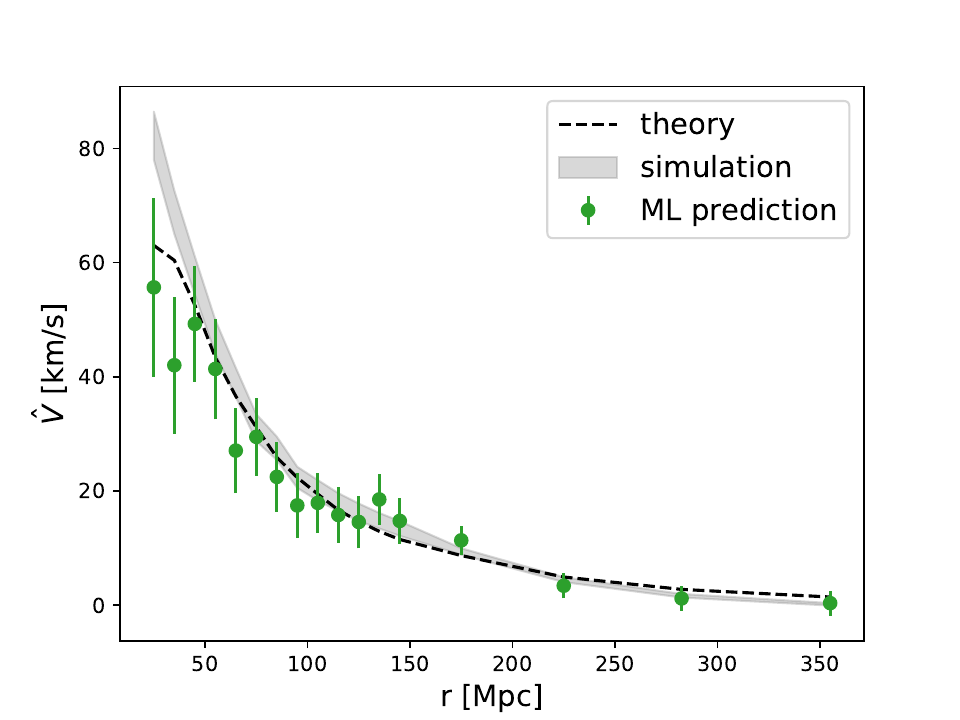}
    \caption{The pairwise velocity statistics and their 1$\sigma$ uncertainties [green circle] for the L60 sample. The results are derived using ILC kSZ temperature measurements and optical depths predicted via the GBDT machine learning algorithm, based on data combining tSZ, $\kappa$, and $M_{vir}$ information for each cluster \cite{Gong:2025gmc}. The predicted linear pairwise velocity for the best-fit cosmology based on \textit{Planck} data is also presented for comparison [black dashed line]. We also present the pairwise velocity derived from the simulation for reference [grey shaded region].}
    \label{fig:Vhat_ML}
\end{figure}

\subsection{Pairwise velocity inference}
\label{sec:Phat}

Using the machine learning method developed in \cite{Gong:2025gmc}, we performed a high-fidelity measurement of the peculiar velocity by estimating the optical depth for each individual galaxy cluster within our L60 sample. We focus specifically on the L60 sample, as its mass range is commensurate with that of the halos in the simulations used to train the machine learning model, ensuring that our measurements closely mirror the conditions under which the model was developed. This minimizes potential biases and allows for a more reliable and accurate application of the model to our observational data.

We employed a multi-probe methodology, extracting synergistic information from different observational tracers of large-scale structure. We extract the features for the machine learning model from the tSZ Compton-$y$ map and the lensing convergence $\kappa$ map, and use virial mass estimates to infer the optical depth. 

The machine learning algorithm was trained to learn the complex, non-linear relationship between these observables and the underlying gas properties, ultimately yielding a high-fidelity estimate of the optical depth for each individual cluster. Combining these machine learning predicted optical depth estimates with the kSZ temperature measurements, we then derive the peculiar velocities of individual clusters using (\ref{eq:vpred}). Using (\ref{eq:Vhat}), we calculate the pairwise velocity from the derived individual galaxy velocities and we estimate the uncertainties with the bootstrap analysis. The scatter and any residual bias in the predicted optical depth, $\tau_{\rm pred}$, are implicitly propagated into the velocity covariance by estimating the covariance of the pairwise velocity using bootstrap resampling of the reconstructed velocities.

The results are presented in Fig.~\ref{fig:Vhat_ML}, along with the linear theoretical prediction based on the best-fit \textit{Planck} cosmology. In the figure, we also present the pairwise velocity derived from the simulation for a sample similar to the L60. As discussed in Sec. \ref{sec:obs}, we apply a correction factor $A_{AP} = 2.73$ to account for the signal attenuation introduced by the aperture photometry measurement \cite{Gong:2025gmc}. This factor is derived by taking the ratio of the pairwise kSZ signal measured with the AP filter to that obtained using a disk filter, following the procedure described in \cite{Gong:2023hse}. We find that the reconstructed pairwise velocity agrees well with the theoretical prediction based on the \textit{Planck} best-fit cosmology for $r > 20$ Mpc, achieving an SNR of 8.5, $\chi^2 = 10$, and PTE = 0.92 for 17 degrees of freedom with 456,803 galaxies. This provides roughly a two-fold improvement in SNR compared to the ACT DR5 × SDSS DR15 analysis reported in \cite{Gong:2025gmc}, which obtained an SNR of 4.5 using 117,384 LRGs. We find that the observed pairwise velocity is consistent with both the linear prediction and the simulation within $1 \sigma$ uncertainty for scales above 40 Mpc.

\section{Conclusion}
\label{sec:conclusion}

In this work, we present a comprehensive analysis of pairwise kSZ curves, using three distinct maps derived from the ACT DR6 data products. Our analysis is based on the f150 and ftot ACT-\textit{Planck} co-added maps, which combine data from both observatories to enhance signal-to-noise ratio, and the component-separated ILC map, which effectively isolates the CMB signal from foreground contaminants. These three maps serve as a robust check on our results, mitigating potential systematic effects inherent to any single data processing pipeline.

We use the LRGs from the DESI DR1 catalog as tracers of the galaxy clusters. We have conducted careful selection of the galaxy (cluster) samples based on their luminosity, a key physical property that correlates with the mass of the host halo while masking out objects located near the galactic plane, close to point sources and massive tSZ clusters, as well as those with high white noise variance in the map. We have divided the DESI catalog into nine distinct luminosity bins. The lowest luminosity cut applied in this study is $L > 3.6 \times 10^{10} L_\odot$. Our careful selection and binning strategy allows us to investigate the dependence of the kSZ signal on cluster properties and to compare our findings with theoretical predictions for the pairwise velocity of the large-scale structure. As discussed in Section~\ref{sec:data}, the inverse-noise cut affects the sample size the most, in addition to the Galactic-plane mask. Adopting a more conservative inverse-noise threshold of 40~$\mu$K/pixel instead of 45~$\mu$K/pixel removes less than 4\% of the sample. The point-source mask and the masking around very massive tSZ clusters have only a minor impact on the sample size. For the Galactic-plane mask, the mask map is available only in three discrete options (30\%, 50\%, and 70\%). Adopting the most conservative 70\% mask would significantly reduce the usable area and hence the sample size. Therefore, the practical freedom to vary this choice is limited. Thus, our main results would be stable under plausible variations of these thresholds. A dedicated quantitative assessment of these effects is beyond the scope of this work and is left for future study.

Using uncertainties estimated from the bootstrap analysis, we derive the mass-averaged optical depth by comparing the observed pairwise kSZ momentum with the theoretical pairwise velocity assuming the best-fit \textit{Planck} cosmology for each sample. Among all the galaxy tracer samples, the highest-significance detection of the pairwise kSZ signal was achieved with the L36 sample using ILC map, which is composed of 913,286 luminous red galaxies. This sample yielded a 9.3$\sigma$ detection, providing a highly significant measurement of the kSZ effect, representing a 72$\%$ increase in the SNR compared with our previous work, which used the ACT DR5 150 GHz CMB map and the SDSS DR15 LRG sample. This represents the highest-significance pairwise kSZ measurement to date, and this high significance underscores the statistical power of our large galaxy catalog. Furthermore, we have obtained consistent measurements of the optical depth when using the co-added f150 and ftot maps, as well as the component-separated ILC map. The fact that our results are consistent across these three independent frequency analyses is a validation of our measurement pipeline. It suggests that potential foregrounds or systematic effects that might vary with frequency do not bias our results, confirming the robustness of our findings. In addition, we also find good agreement between the results obtained from the full sample and those from the low-redshift subsample.

To further validate the robustness of our kSZ measurements and its derived fitted mass-averaged optical depth, we performed a simple cross-check by estimating the optical depth from the simulation for each sample. This estimation was also derived by comparing the pairwise kSZ curves with the pairwise velocities measured from the simulation. This simulation-derived optical depth serves as an independent reference point for comparison with our primary kSZ-derived results. We found that the optical depth estimated from this simple method agrees well with the ACT $\times$ DESI optical depth inferred from our pairwise kSZ measurements, particularly those derived from both the 150 GHz and the ILC maps. The consistency of these values, well within the 1$\sigma$ uncertainty, provides strong support for the reliability of our kSZ-based methodology and the derived mass-averaged optical depth values. We emphasize that a more detailed study of the tSZ-based estimation will be presented in a future companion paper.

Finally, we implemented a machine learning model to estimate the optical depth for each individual galaxy in our sample, moving beyond the mass-averaged values. The model was trained on a rich set of features, including tSZ measurements, lensing convergence $\kappa$, and cluster mass estimates. By combining these estimated individual optical depths with the measured kSZ signal, we were able to infer the peculiar velocity of each cluster. From these individual velocities, we then computed the pairwise velocity statistic, which probes the dynamics of the large-scale structure. Analyzing the substantial L60 sub-sample, of 456,803 galaxies, we achieved a significant 8.5$\sigma$ detection of the pairwise velocity. The measured pairwise velocity is found to be in excellent agreement with the linear theoretical predictions derived from the best-fit \textit{Planck} cosmology, and consistent with the simulation predictions at scales above 40 Mpc.

This comprehensive analysis, which has successfully demonstrated a robust pipeline for measuring the pairwise kSZ effect, establishes a strong foundation for future studies leveraging ACT data with forthcoming DESI data releases. The next generation of cosmic microwave background (CMB) experiments and large-scale structure surveys will further enhance the potential of kSZ science. Upcoming CMB observatories, such as the Simons Observatory \cite{SimonsObservatory:2018koc} and CCAT \cite{CCAT-Prime:2021lly}, are designed to deliver high-resolution, low-noise CMB maps over unprecedentedly large areas of the sky. These maps will be essential for resolving the faint kSZ signal with high fidelity. Concurrently, a new wave of large-scale structure surveys, including DESI DR2, Euclid \cite{Euclid:2024yrr}, and Roman \cite{spergel2015widefieldinfrarredsurveytelescopeastrophysics}, along with the Rubin telescope \cite{10.71929/rubin/2570308} will provide complementary optical and infrared datasets with exquisite photometric and spectroscopic coverage. The combination of these efforts will enable cross-correlation studies with unparalleled statistical power, pushing kSZ measurements beyond their current limits. This will make the kSZ effect, and its inferences about the cosmic velocity fields, a competitive and independent probe of cosmology, providing crucial insights into the nature of dark energy and the validity of General Relativity on cosmological scales.

\begin{acknowledgments}
 The work of YG and RB is supported by NSF grant AST-2206088,  NASA grant 22-ROMAN11-0011, and NASA grant 12-EUCLID12-0004. CS acknowledges support from the Agencia Nacional de Investigaci\'on y Desarrollo (ANID) through Basal project FB210003.

Support for ACT was through the U.S.~National Science Foundation through awards AST-0408698, AST-0965625, and AST-1440226 for the ACT project, as well as awards PHY-0355328, PHY-0855887 and PHY-1214379. Funding was also provided by Princeton University, the University of Pennsylvania, and a Canada Foundation for Innovation (CFI) award to UBC. ACT operated in the Parque Astron\'omico Atacama in northern Chile under the auspices of the Agencia Nacional de Investigaci\'on y Desarrollo (ANID). The development of multichroic detectors and lenses was supported by NASA grants NNX13AE56G and NNX14AB58G. Detector research at NIST was supported by the NIST Innovations in Measurement Science program. Computing for ACT was performed using the Princeton Research Computing resources at Princeton University, the National Energy Research Scientific Computing Center (NERSC), and the Niagara supercomputer at the SciNet HPC Consortium. SciNet is funded by the CFI under the auspices of Compute Canada, the Government of Ontario, the Ontario Research Fund–Research Excellence, and the University of Toronto. We thank the Republic of Chile for hosting ACT in the northern Atacama, and the local indigenous Licanantay communities whom we follow in observing and learning from the night sky.

This research used resources of the National Energy Research Scientific Computing Center (NERSC), a U.S. Department of Energy Office of Science User Facility located at Lawrence Berkeley National Laboratory, operated under Contract No. DE-AC02-05CH11231 using NERSC award HEP-ERCAPmp107

This material is based upon work supported by the U.S. Department of Energy (DOE), Office of Science, Office of High-Energy Physics, under Contract No. DE–AC02–05CH11231, and by the National Energy Research Scientific Computing Center, a DOE Office of Science User Facility under the same contract. Additional support for DESI was provided by the U.S. National Science Foundation (NSF), Division of Astronomical Sciences under Contract No. AST-0950945 to the NSF’s National Optical-Infrared Astronomy Research Laboratory; the Science and Technology Facilities Council of the United Kingdom; the Gordon and Betty Moore Foundation; the Heising-Simons Foundation; the French Alternative Energies and Atomic Energy Commission (CEA); the National Council of Humanities, Science and Technology of Mexico (CONAHCYT); the Ministry of Science, Innovation and Universities of Spain (MICIU/AEI/10.13039/501100011033), and by the DESI Member Institutions: \url{https://www.desi.lbl.gov/collaborating-institutions}. Any opinions, findings, and conclusions or recommendations expressed in this material are those of the author(s) and do not necessarily reflect the views of the U. S. National Science Foundation, the U. S. Department of Energy, or any of the listed funding agencies.

The authors are honored to be permitted to conduct scientific research on I'oligam Du'ag (Kitt Peak), a mountain with particular significance to the Tohono O’odham Nation.

\end{acknowledgments}

\section*{DATA AVAILABILITY}
The data corresponding to the figures in this paper are available at https://doi.org/10.5281/zenodo.17373480.

A detailed description of the pipeline used for aperture photometry and pairwise computation can be found in \cite{Gallardo:2025lzu}.

\appendix
\section{Null test}
\label{sec:appendix_a}

\begin{figure}
    \centering
    \includegraphics[width=\columnwidth]{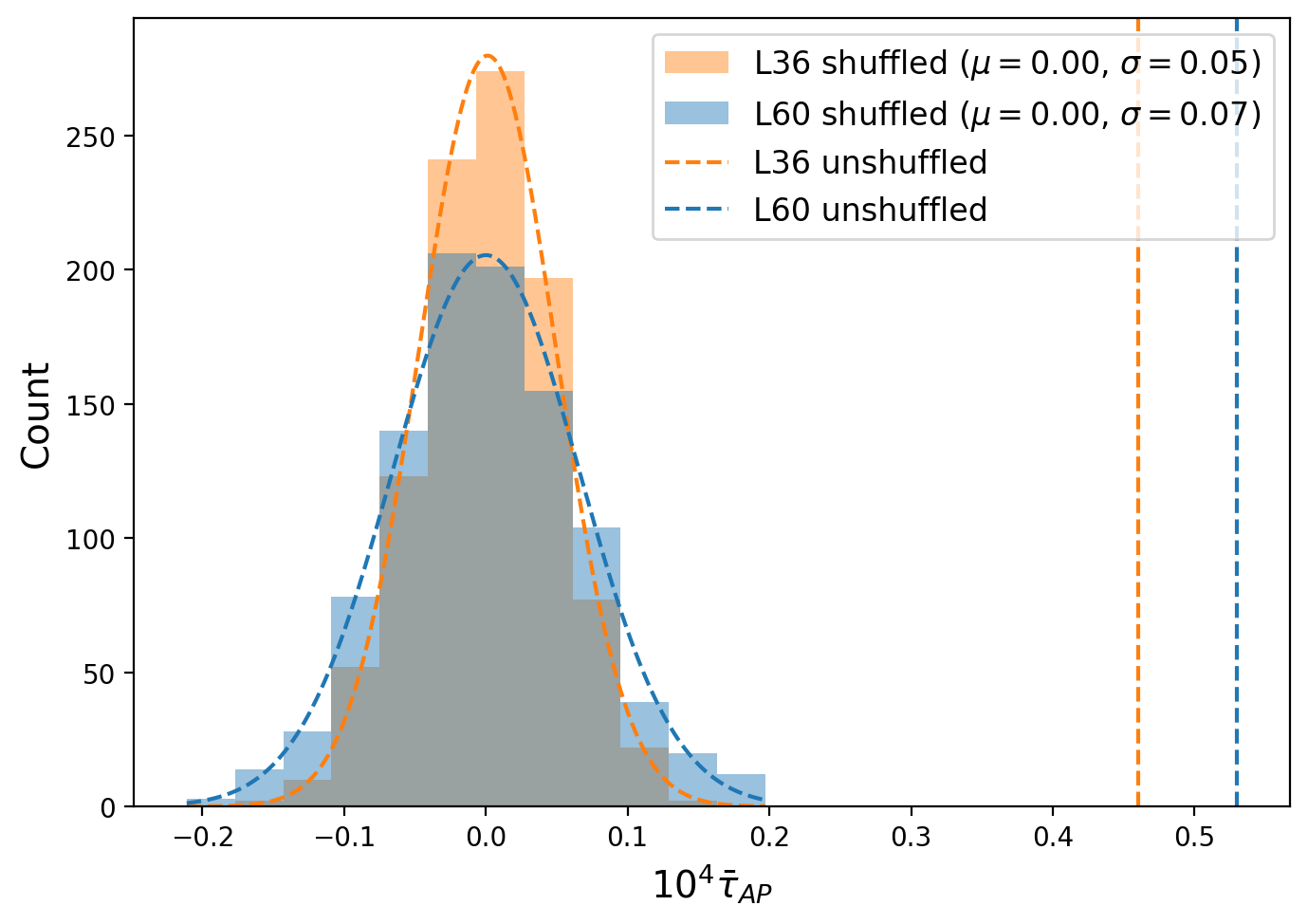}
    \caption{Histogram of the best-fit $\bar{\tau}$ values from 1,000 null-test realizations for the L36 [orange] and L60 [blue] sample, compared with the theoretical pairwise velocity prediction derived from the best-fit \textit{Planck} cosmology. The curve represents the fitted normal distribution, with mean zero and a standard deviation of 0.05 and 0.07 respectively. The best-fit $\bar{\tau}$ value of the unshuffled result for the L36 and L60 sample is shown as the vertical dashed line.}
    \label{fig:nulltest}
\end{figure}
To ensure the robustness and validity of our pairwise kSZ measurements, we conducted a null test. The null test serves as a critical verification step, assessing the possibility of systematic errors or biases inherent in the dataset or the analysis methodology.

In practice, the null test involves randomly shuffling galaxy positions while preserving their redshift distribution. Through this randomized procedure, we generate synthetic galaxy catalogs that maintain the statistical characteristics of the original dataset but lack genuine physical correlations between galaxy pairs.

Applying the pairwise kSZ estimator to randomized datasets is expected to yield no significant signal, thereby establishing a baseline for the absence of physical correlations. The results of this null test conducted on the L36 and L60 sample using the ILC map are shown in Figure \ref{fig:nulltest}, where the best-fit $\bar\tau$ values derived from the pairwise kSZ signal measured from 1,000 shuffled realizations clusters tightly revolve around zero, consistent with expectations. The fitted optical depth $\bar{\tau}$ also follows a Gaussian distribution with mean zero. This
validates the $\sim 9 \sigma$ and $\sim 6 \sigma$ significance of the unshuffled result shown as the vertical dashed line in the figure.

The consistency of our null test results with zero velocity confirms that our pairwise kSZ measurements are free from significant systematic biases. Any observed deviations from zero in the actual data can therefore be confidently interpreted as genuine physical signals indicative of coherent galaxy motions rather than artifacts from observational or analytical procedures.

\clearpage
\bibliography{draft}

\end{document}